\documentclass[apj]{emulateapj}
\usepackage{hyperref}
\shorttitle{Recurrent $^{3}$He-Rich SEPs}
\shortauthors{R.~Bu\v{c}\'ik et al.}

\begin{document}


\title{Multi-Spacecraft Observations of Recurrent $^{3}$He-Rich Solar
Energetic Particles}


\author{R.~Bu\v{c}\'ik and D.~E.~Innes and U.~Mall and A.~Korth}
\affil{Max-Planck-Institut f\"{u}r Sonnensystemforschung, D-37191 Katlenburg-Lindau\altaffilmark{1}, Germany}
\email{bucik@mps.mpg.de}

\author{G.~M.~Mason}
\affil{Applied Physics Laboratory, Johns Hopkins University, Laurel, MD 20723, USA}
%\email{aastex-help@aas.org}

\and

\author{R.~G\'omez-Herrero}
\affil{Space Research Group, University of Alcal\'a, E-28871 Alcal\'a de Henares, Spain}

\altaffiltext{1}{since 2014 February: D-37077 G\"{o}ttingen}

\begin{abstract}
We study the origin of $^{3}$He-rich solar energetic
particles ($<$1~MeV\,nucleon$^{-1}$) that are observed consecutively on {\sl STEREO-B},
{\sl ACE}, and {\sl STEREO-A} spacecraft when they are separated in heliolongitude by
more than 90\degr. The $^{3}$He-rich period on {\sl STEREO-B} and {\sl STEREO-A} commences
on 2011 July 1 and 2011 July 16, respectively. The {\sl ACE} $^{3}$He-rich
period consists of two sub-events starting on 2011 July 7 and 2011 July 9.
We associate the {\sl STEREO-B} July 1 and {\sl ACE} July 7 $^{3}$He-rich
events with the same sizeable active region producing X-ray flares accompanied by
prompt electron events, when it was near the west solar limb as seen from the
respective spacecraft. The {\sl ACE} July 9
and {\sl STEREO-A} July 16 events were dispersionless with enormous $^{3}$He enrichment,
lacking solar energetic electrons and occurring in corotating interaction
regions. We associate these events with a small, recently emerged active region
near the border of a low-latitude coronal hole that produced numerous jet-like emissions temporally
correlated with type III radio bursts. For the first time we present
observations of 1) solar regions with long-lasting conditions for $^{3}$He
acceleration and 2) solar energetic $^{3}$He that is temporary confined/re-accelerated
in interplanetary space.
\end{abstract}





\section{INTRODUCTION}

Solar energetic particle (SEP) $^{3}$He-rich events are characterized by huge
abundance enhancements of the rare isotope $^{3}$He (up to factors of
$>$10$^{4}$) over solar system abundances.
The enrichment of the $^{3}$He is believed to be caused by selective heating due to
wave-particle interaction in the flare plasma because of its unique charge to
mass ratio \citep[see e.g., reviews by][]{koc84,rea90}. Earlier studies have
shown that $^{3}$He-rich SEP events are associated with 2-100~keV electron events
\citep{rea85} and their related type III radio bursts \citep{rea86}, which are
excited by electrons streaming outward through interplanetary space \citep[e.g.,][]{lin74}.
The events have been associated with minor soft X-ray flares
\citep[e.g.,][]{zwi78,kah87,rea88}. Recently the sources
of $^{3}$He-rich events have been investigated using extreme ultraviolet (EUV)
and X-ray solar images \citep{nit06,nit08,wan06}. The responsible source has
often been attributed to a jet-like ejection near a coronal hole.

Sequences of $^{3}$He-rich SEP events from the same active region (AR) have been
observed with a single spacecraft (s/c) \citep{rea86,mas99,mas00,wan06,pic06}
only in a limited time interval (about one day) presumably due to loss of the
magnetic connection to the flare site. These recurrent events may suggest almost
continuous acceleration of $^{3}$He in the single AR \citep{pic06}.
In addition, multi-day periods of solar energetic $^{3}$He have been recently
discovered with the {\sl Advanced Composition Explorer} ({\sl ACE}) often with no individual
$^{3}$He-rich events resolved \citep[and references therein]{mas07}. Such long periods
have been interpreted as a combination of continuous production of $^{3}$He-rich SEPs
and their confinement in interplanetary magnetic field structures \citep{koc08}.

In this paper we report, for the first time, on a $^{3}$He-rich period of SEPs
consecutively observed by three widely, in longitude, separated spacecraft
{\sl STEREO-B}, {\sl ACE} and {\sl STEREO-A} with delays corresponding to the Carrington
rotation rate. The ultimate goal of this study is to identify activity on
the Sun and the conditions in interplanetary space leading to the long
persistence of energetic $^{3}$He in the heliosphere. In Section~\ref{epo}, we
describe energetic particle and solar wind plasma observations on each s/c.
The responsible solar sources are identified in Section~\ref{ss} using a combination of
coronal field model extrapolations, EUV and radio observations. The results
are summarized and discussed in Section~\ref{sad}.

\section{ENERGETIC PARTICLE OBSERVATIONS} \label{epo}

%% In a manner similar to \objectname authors can provide links to dataset
%% hosted at participating data centers via the \dataset{} command.  The
%% second curly bracket argument is printed in the text while the first
%% parentheses argument serves as the valid data set identifier.  Large
%% lists of data set are best provided in a table (see Table 3 for an example).
%% Valid data set identifiers should be obtained from the data center that
%% is currently hosting the data.
%%
%% Note that AASTeX interprets everything between the curly braces in the
%% macro as regular text, so any special characters, e.g. "#" or "_," must be
%% preceded by a backslash. Otherwise, you will get a LaTeX error when you
%% compile your manuscript.  Special characters do not
%% need to be escaped in the optional, square-bracket argument.

The helium isotope observations described in this paper are from the time-of-flight
mass spectrometers Suprathermal Ion Telescope \citep[SIT;][]{mas08}
on the two {\sl Solar Terrestrial Relations Observatories} ({\sl STEREO}) and Ultra Low
Energy Isotope Spectrometer \citep[ULEIS;][]{mas98} on {\sl ACE}. The instruments
have sunward viewing directions close to the average Parker magnetic field
spiral line ({\sl STEREO}) or in the sunward hemisphere ({\sl ACE}). They measure ions
from He to Fe in the energy range from 20~keV\,nucleon$^{-1}$ to several
MeV\,nucleon$^{-1}$. Both {\sl STEREO} s/c
are in a heliocentric orbit at $\sim$1~AU near the ecliptic plane increasing
their separation from Earth at a rate of $\sim$22\degr\ per year. {\sl STEREO-A} is preceding Earth and
{\sl STEREO-B} is trailing behind. The {\sl ACE} s/c is in an orbit around the L1
Lagrangian point, $\sim$0.01~AU upstream of Earth in the sunward direction.
During the $^{3}$He-rich period investigated in this paper {\sl STEREO-B} was 93\degr\ behind
and {\sl STEREO-A} 99\degr\ ahead of the Earth with the heliocentric distances of 1.02
and 0.96~AU, respectively. These angular separations correspond to corotation
delays of 7.1 and 7.5 days, respectively, using the Carrington rotation
period of 27.3 days.

Since the magnetic connection of the spacecraft to the Sun lies west of the
central meridian viewed by each spacecraft, the {\sl GOES} spacecraft at Earth was
able to observe source regions for {\sl STEREO-B} as well as {\sl ACE}, but not for {\sl STEREO-A}.

\subsection{STEREO-B $^{3}$He-Rich Period} \label{bp}

\begin{figure}
%\epsscale{.6}
\plotone{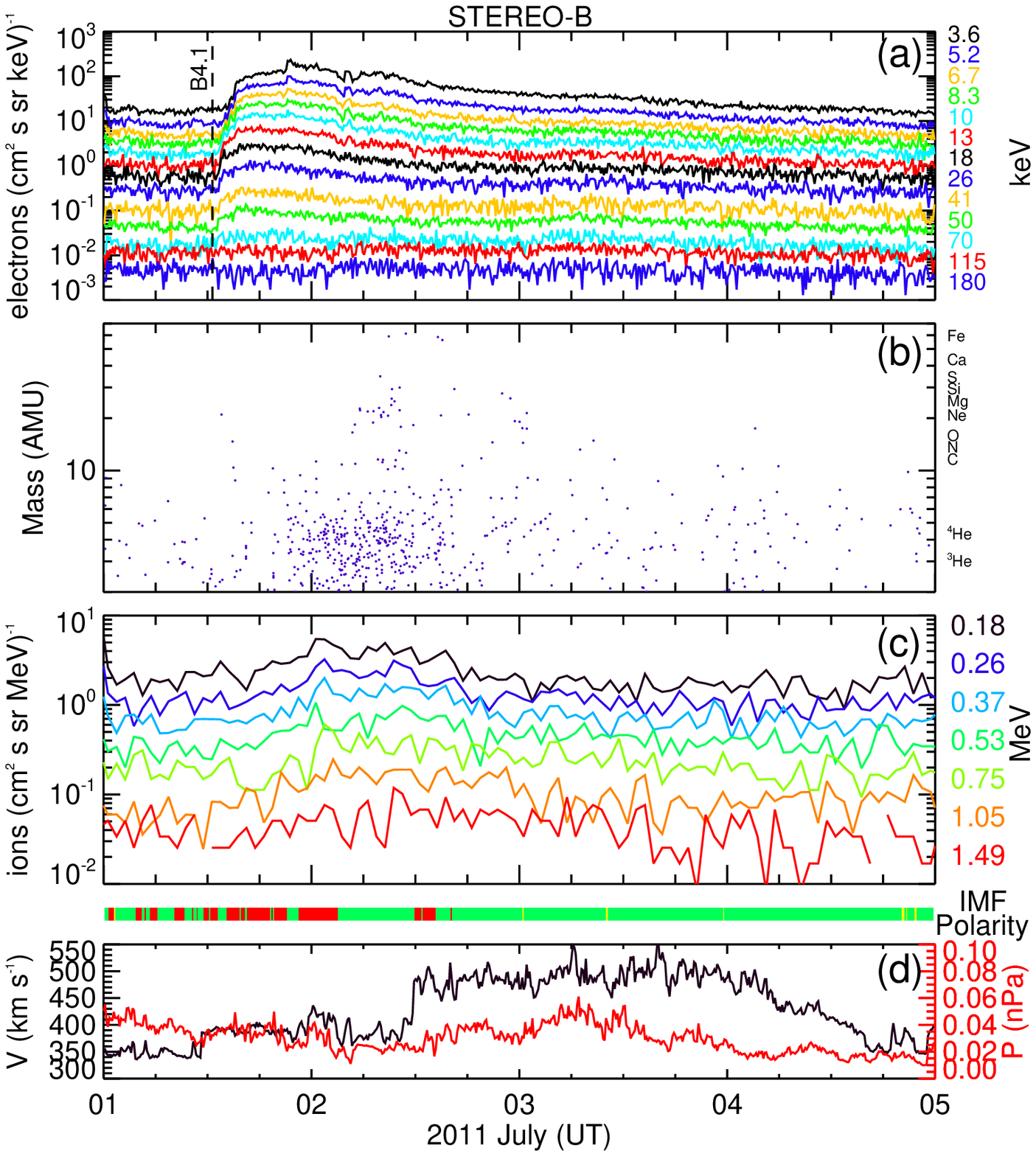}
\caption{(a) 10-min {\sl STEREO-B}/STE D3 (3.6-41~keV) electron intensities
from one of the four downstream (looking in anti-sun direction) STE
detectors and 10-min {\sl STEREO-B}/SEPT (50-180~keV) electron intensities
from the sunward pointing sensor. The dashed vertical line indicates the start time of the {\sl GOES}
B4.1 X-ray flare. (b) {\sl STEREO-B}/SIT mass spectrogram of individual ions in
the energy ranges 0.25-0.90~MeV\,nucleon$^{-1}$ (mass $<$8~amu) and
0.08-0.15~MeV\,nucleon$^{-1}$ (mass $>$8~amu). (c) 1-hr {\sl STEREO-B}/SEPT
(0.18-1.49~MeV) ion intensities from the sunward pointing sensor. (d) 10-min solar
wind speed $V$ (black curve) and 10-min total pressure $P$ (red curve).
IMF polarity color bar indicates by red
(green) the times when the observed magnetic field vector was oriented
toward (away from) the Sun. The yellow indicates the times with ambiguous
polarity.  \label{fig1}}
\end{figure}

Figure~\ref{fig1} presents energetic particle and solar wind plasma measurements for
the $^{3}$He-rich period starting on 2011 July 1 on {\sl STEREO-B}. Figure~\ref{fig1}a
shows {\sl STEREO-B} 10-min electron intensities from the STE-D \citep{lin08}
anti-sunward pointing detector and from the SEPT \citep{mul08}
sunward pointing sensor centered on the nominal Parker spiral direction.
The data from the sunward pointing STE-U are not available, because the instrument
is saturated by sunlight \citep{wan12}. A solar energetic electron event
was clearly observed on 2011 July 1 with an onset at $\sim$13:10~UT and a velocity
dispersion down to 3~keV. At higher energies ($>$100~keV) increases in the electron
intensity were not detected by SEPT. The event was associated with
B4.1 {\sl GOES} X-ray flare in the 1-8~{\AA} range and H$\alpha$ flare with a start
time at 12:34~UT. According to the Edited Solar Events Lists\footnote{\url{www.swpc.noaa.gov}}
compiled by the NOAA Space Weather Prediction Center (SWPC) the flare originated
from AR 11244 (N15\degr W05\degr\ as viewed from Earth) and was accompanied
by a type III radio burst. From the {\sl STEREO-B} point of view the AR was located at
$\sim$W97\degr.

\begin{figure}
%\epsscale{.7}
\plotone{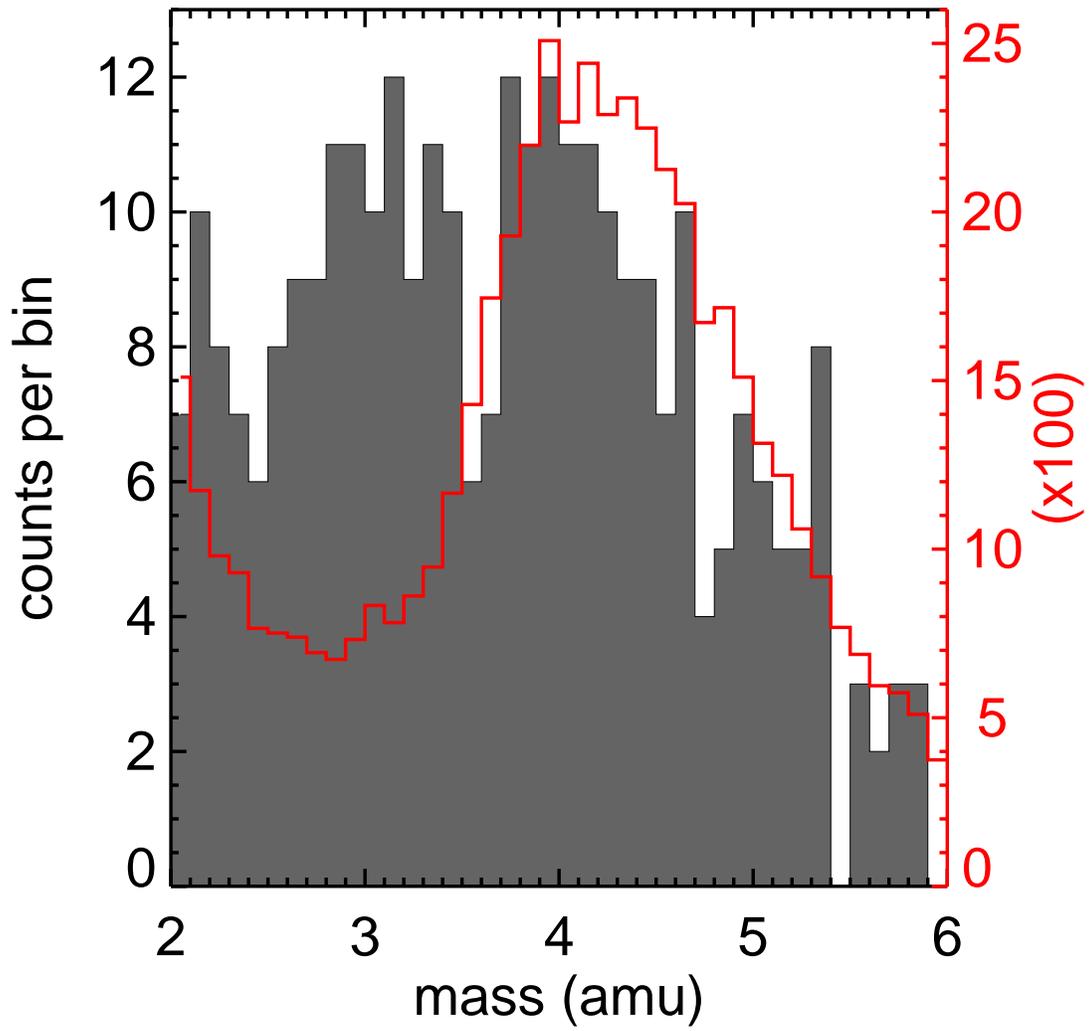}
\caption{{\sl STEREO-B}/SIT He mass histograms in the energy range 0.25-0.9~MeV\,nucleon$^{-1}$
for 2011 July 1 $^{3}$He-rich period (gray shaded) and for an intense CIR event
on 2010 August 19-21 (red curve). \label{fig2}}
\end{figure}

The period of $^{3}$He-rich SEPs on 2011 July 1 shows quite low ion increases. Figure~\ref{fig1}b
shows the SIT mass spectrogram of all individual ions in the energy
ranges 0.25-0.90~MeV\,nucleon$^{-1}$ (mass $<$8~amu) and 0.08-0.15~MeV\,nucleon$^{-1}$
(mass $>$8~amu). The spectrogram indicates that the number density of the ions
started to increase roughly around 21~UT. Taking the lower energy limit of
the ions in Figure~\ref{fig1}b the estimated injection on the Sun would be at
14~UT. The 1-hr SEPT ion intensities in Figure~\ref{fig1}c show only a minor
(factor of $\sim$3) enhancement above the background. Observations in
Figure~\ref{fig1}b show that the SIT pulse-height points are spread over
the whole helium mass range. To rule out $^{4}$He spillover to the $^{3}$He we compare
in Figure~\ref{fig2} helium SIT-B histograms for the 2011 July 1 period and a typical
corotating interaction region (CIR) event, which is almost all $^{4}$He. The figure
clearly indicates a separate $^{3}$He peak in the valley of the CIR histogram.
The $^{3}$He/$^{4}$He elemental ratio was determined by summing the counts in the two
clearly separated He isotope peaks in the mass histogram. Note that at higher
energies the \textsl{STEREO-B}/LET \citep{mew08} instrument shows no $^{3}$He
increase on July 1-2 in the He mass spectrogram plots in intervals
2.3-3.3, 4.3-8.0~MeV\,nucleon$^{-1}$. The 385~keV\,nucleon$^{-1}$
$^{3}$He/$^{4}$He and Fe/O abundance ratios for the $^{3}$He-rich period in
this study on all three spacecraft are
listed in Table~\ref{tab1}. The table further shows s/c Carrington longitudes and
heliographic latitudes, the start and end times of the period on the different spacecraft,
and the $^{3}$He fluences.

Figure~\ref{fig1}d shows 10-min averages of solar wind speed $V$ and the total
pressure $P$, which is given by the sum of the thermal and magnetic pressure,
i.e., $P \sim n_{p}k(1.16T_{p}+1.3\times10^{5}) + B^{2}/2\mu_{0}$, where $n_{p}$
and $T_{p}$ are the proton density and temperature, respectively, and $B$ is
the magnetic field magnitude. The thermal pressure, taken from \citet{jia06},
includes the contribution of solar wind electrons and alpha particles. Solar
wind proton and magnetic field data were obtained from the PLASTIC \citep{gal08}
and MAG \citep{acu08} instruments on {\sl STEREO}. The total pressure shows no
increase over the ambient solar wind pressure. The onset of a high-speed solar
wind on 2011 July 2 was preceded by the magnetic field polarity reversal. IMF
(interplanetary magnetic field) polarity color bar in Figure~\ref{fig1} indicates in red
(green) the times when the observed magnetic field vector was oriented
toward (away from) the Sun. The polarity sectors are defined as $\pm$70\degr\
about the nominal Parker angle. The Parker angle was calculated using the observed solar
wind speed. The yellow indicates when the magnetic field vector was within the range $\pm$20\degr\
about the normal to the nominal Parker magnetic field direction.

\subsection{ACE $^{3}$He-Rich Period} \label{l1p}

Figure~\ref{fig3} presents energetic particle and solar wind plasma parameters
throughout the {\sl ACE} $^{3}$He-rich period starting on 2011 July 7 and terminating at
the end of 2011 July 10. Figure~\ref{fig3}a shows 5-min energetic electron intensities.
The near-relativistic electrons are routinely monitored on {\sl ACE}, but because
of the data gap between 05 and 11~UT on July 8 the intensities are shown from
the {\sl WIND}/3DP \citep{lin95}, located also near the L1 point. The measurements in Figure~\ref{fig3}a
clearly show four impulsive solar energetic electron events with onset times
at $\sim$ 05:50, 14:50, 20:55~UT on 2011 July 7 and 03:20~UT on 2011 July 8 at
182~keV. On shorter time scales, velocity dispersion is evident in all
these electron events. The events were associated with B2.6, B7.6, B6.4, B3.0
X-ray flares with start times at 05:09, 14:25, 20:24~UT on July 7
and at 02:53~UT on July 8, respectively. According to the NOAA SWPC
list the B2.6 and B3.0 flares originated from AR 11244 (N15\degr W82\degr) and AR 11243
(N16\degr W60\degr), respectively. On 2011 July 7 both solar regions were associated
with sunspots; AR 11243 contained several bipolar sunspots and AR 11244 only
one visible unipolar sunspot as reported in the NOAA Solar Region Summary (SRS). The B7.6 and B6.4
flares do not have an assigned solar region in the event list. All these four
flares were accompanied by type III solar radio bursts detected by different
ground-based observatories, as reported in the NOAA list. No related H$\alpha$ flare
is recorded in the H-Alpha Solar Flare Report at the NOAA National Geophysical
Data Center\footnote{\url{www.ngdc.noaa.gov}}. The arrow in Figure~\ref{fig3}a
indicates the time of the EUV jet on July 8 (16:25~UT) identified in the
next section (Section~\ref{9ev}).

The 1-hr 0.23-0.32~MeV\,nucleon$^{-1}$ $^{3}$He intensity in Figure~\ref{fig3}b shows two separate events
or sub-periods, one starting near the end of 2011 July 7 ($\sim$21:00~UT) and another
at the beginning of 2011 July 9 ($\sim$01:00~UT). Two double-ended arrows in
Figure~\ref{fig3}b denote the {\sl STEREO-B} and {\sl STEREO-A} $^{3}$He-rich periods (as in Table \ref{tab1})
shifted in corotation time. The partial overlap of the periods on all three s/c
suggests a common solar origin. Due to its smaller geometrical factor, the SIT
instrument is less sensitive to the weak $^{3}$He-rich events than the ULEIS.
Thus the $^{3}$He-rich periods observed by ULEIS have generally longer duration.
The high energy (0.4-10~MeV\,nucleon$^{-1}$) mass spectrogram data (Figure~\ref{fig3}c)
show that the first $^{3}$He ions arrived around 12:00~UT on 2011 July 7. These are likely
to be related to the first electron event seen on the same day. The 2011 July 7
$^{3}$He-rich event itself may have come from multiple injections. The beginning
of this period was probably associated with the second electron event. The estimated
delay between 100~keV electrons and 0.23-0.32~MeV\,nucleon$^{-1}$ ions traveling along
the same path length from the Sun to L1 would be $\sim$6.5 hours, which approximately
agrees with the observed time offset. There is a small additional $^{3}$He intensity increase
observed at the beginning of 2011 July 8 followed by increases in the heavy-ion
(Fe and O) intensities (see Figure~\ref{fig3}b). This increase could be
related to the most intense electron event seen at the end of 2011 July 7. It is not
clear if any $^{3}$He was associated with the solar electron event at the beginning
of July 8. The presence of an enhanced $^{3}$He intensity from the previous activity
along with a data gap at 05-11~UT would obscure any $^{3}$He injection associated
with this electron event. There were no observed electron increases closely
preceding the 2011 July 9 $^{3}$He-rich event. A small dispersed electron
event at energy $<$10~keV is seen on July 9 after the SEP event
start time. From Figure~\ref{fig3}b we can see that
the July 9 event is more enriched in $^{3}$He and also shows higher $^{3}$He intensities
compared to the event on July 7. The Fe and O time lines lie nearly on
top of each other for both events. The abundance ratios for these two {\sl
ACE} events are listed in Table~\ref{tab1}. The time intervals used for calculating the ratios
and fluences are only approximate and roughly coincide with the enhanced Fe/O
ratio. The {\sl ACE}/SIS \citep{sto98} instrument in the energy intervals
4.5-7.6, 7.6-16.3~MeV\,nucleon$^{-1}$ shows a high $^{3}$He/$^{4}$He ratio
($\sim$0.5-1.0) for the July 7 period in the spectrogram plots at the ACE science
center\footnote{\url{www.srl.caltech.edu/ACE/ASC/DATA/level3/sis/heplots/}}.
On July 9 the $^{3}$He in these high energy ranges is only seen during
00-06~UT. It was probably a continuation from previous day and related to the July 7
event. Figure~\ref{fig3}d shows 1-hr {\sl ACE}/EPAM \citep{gol98}
ion intensities at 120\degr\ from the sunward pointing axis. The data from the telescope looking towards
the Sun (LEMS30) are not available for energies $<$1~MeV. The intensity-time
profiles in Figure~\ref{fig3}d indicate a velocity dispersion at the onset of the July 7
event and a dispersionless onset for the July 9 event.

The IMF polarity color bar in Figure~\ref{fig3} indicates
that the two $^{3}$He-rich events occurred in opposite magnetic polarity sectors.
In addition, the intensity plots in Figures~\ref{fig3}b and \ref{fig3}d show
that the end of July 7 event and the start of July 9 event coincide with the
interplanetary magnetic sector boundary crossing. The solar wind data in Figure~\ref{fig3}e
reveal a CIR on July 9, characterized by a gradual solar wind speed rise and
the pressure enhancement. Because of the data gap on {\sl ACE} the proton plasma density
used in the total pressure is from the {\sl WIND}/SWE \citep{ogi95} instrument.
The remaining solar wind parameters are from MAG \citep{smi98} and SWEPAM
\citep{mcc98} on {\sl ACE}. CIRs arise when fast solar wind, emerging from coronal
holes, overtakes the preceding slow solar wind. The leading and trailing edge of
the interaction region are characterized by an increase and decrease in the total
pressure, respectively, and the enhanced pressure occurs in the rising part of
the speed profile \citep[e.g.,][]{bur95}. There is a gradual change in the flow
direction from near radial to the east (refer to Figure~\ref{fig3}f) at the beginning
of July 9, which is synchronized with the start of the slow rise in the solar
wind speed. Although the boundary between the two solar wind regimes here is
not sharp, such a transition can indicate an evolving stream interface. Apart from
pure time evolution the s/c heliographic latitude could also be playing some
role in the observed high-speed stream structure. We notice from Figure~\ref{fig3} that
the July 9 $^{3}$He-rich event commenced before the interface, and close
to the time that the change of the magnetic polarity occurred.
It is interesting that the period of relatively enhanced
$^{3}$He on July 10 coincides with large variations in the high-speed solar wind
and further deflection of the flow toward the east.

\subsection{STEREO-A $^{3}$He-Rich Period} \label{ap}

\subsubsection{General Properties} \label{gp}

The $^{3}$He-rich period of interest starts for {\sl STEREO-A} on 2011 July 16. The simultaneous
energetic particle and solar wind plasma measurements during the $^{3}$He-rich
period on {\sl STEREO-A} are presented in Figure~\ref{fig4}. Figure~\ref{fig4}a shows 10-min electron
intensities from the STE-D anti-sunward and the SEPT sunward pointing sensors. No obvious
solar electron event is seen in the presented time range. On 2011 July 16
the electron intensities were near background level. The STE electron intensities
at higher energies were dominated by energetic ions (see \citet{wan12} for a description
of the background in STE). Arrows in Figure~\ref{fig4}a
denote EUV brightenings (12:40, 14:15, 18:45~UT) and a jet (22:25~UT) on July 15
identified in the next section (Section~\ref{aev}). A marked $^{3}$He enhancement
is seen in the SIT mass spectrogram in Figure~\ref{fig4}b. There is a data gap on
July 16 from 09:45 to 11:57~UT in the SIT measurements at the beginning of
the $^{3}$He-rich period. Therefore we plot in Figure~\ref{fig4}c 1-hr ion intensities
between 0.18 and 1.49~MeV from the SEPT sunward sensor which did not suffer
from this data gap. The onset at $\sim$09:30~UT is clearly dispersionless with a simultaneous
rise of ion intensities in all energy channels. Note that the approximate delay
time between ions in the lowest and highest energy channels in Figure~\ref{fig4}c would
be at least 5 hours. This estimate is based on scatter-free ion propagation
(with zero pitch angle) along the spiral with a path length of 1.1~AU to {\sl
STEREO-A}, assuming a solar wind speed of 400~km\,s$^{-1}$. Dispersionless onset indicates that
the first solar energetic particles arrived to 1~AU before the s/c enters the flare
connected field lines \citep{koc08}. This could also be the reason
for the lack of electron enhancement. Dispersionless $^{3}$He-rich events associated
with interplanetary shocks or magnetic clouds were reported by \citet{tsu02}.
The SEPT ion enhancements shown in Figure~\ref{fig4}c are relatively minor and comparable
with the pre-event intensities. The statistical errors of SEPT ion intensities
during the $^{3}$He-rich period are low and therefore not shown. For instance,
the errors range between 0.1 and 0.3 in the highest energy channel.
The second (lower) black curve in Figure~\ref{fig4}c shows the 1.49~MeV ion intensity from
the SEPT anti-sunward pointing detector. The energetic ion fluxes appear to be
only slightly anisotropic with intensities in the sunward pointing detector higher
by about a factor of $\sim$2. Similar small anisotropies were seen in the lower energy
channels. The $^{3}$He-rich events usually exhibit large (outward to sunward ratio $>$10)
and long lasting anisotropies~\citep[e.g.,][]{zwi78}. We remark that SEPT
is not capable of elemental resolution and most of the time protons are
the contributors in the ion channels. However, for a $^{3}$He-rich
SEP event, helium can account for a sizable amount \citep{mul08}.
The elemental abundances for 2011 July 16 $^{3}$He-rich event have been examined in an earlier report by
\citet{buc13b}. This study shows that heavy ions are also enriched in
this event with the relative abundances of elements (Ne-S) and Fe typical
of $^{3}$He-rich SEP events \citep[e.g.,][]{mas04}. The 385~keV\,nucleon$^{-1}$
$^{3}$He/$^{4}$He and Fe/O ratios for this event are in Table~\ref{tab1}.
The {\sl STEREO-A}/LET instrument shows a clear $^{3}$He increase for the July 16 event
in the energy interval 2.3-3.3~MeV\,nucleon$^{-1}$ with a high $^{3}$He/$^{4}$He ratio ($>$1)
but not in 4.3-8.0~MeV\,nucleon$^{-1}$.

Figure~\ref{fig4}d shows 10-min averages of solar wind speed $V$ and total pressure $P$.
The shaded region marks the time interval of a CIR obtained from a
list\footnote{\url{www-ssc.igpp.ucla.edu/forms/stereo/stereo\_level\_3.html}} compiled
by the {\sl STEREO} magnetometer team at the University of California Los
Angeles (UCLA), based on plasma and magnetic field data \citep{jia06}. Note
that the total pressure $P$ peak is a factor of $\sim$2 higher than in
the CIR on 2011 July 9. We see in Figure~\ref{fig4}d that the CIR trailing boundary is
well-defined with a jump-like decrease (increase) in the pressure $P$ (speed $V$),
suggestive of a developing reverse shock. The dashed vertical line indicates
the stream interface (SI), separating slow and fast stream plasma, characterized
by a sharp discontinuity in the azimuthal flow direction (Figure~\ref{fig4}e) and a jump
in the flow speed (Figure~\ref{fig4}d) \citep{gos78}. Other characteristics
of the stream interface such as abrupt and simultaneous drops in plasma proton
density and rise in proton temperature were also observed. A more detailed
investigation of the solar wind plasma data revealed a smooth North to South
rotation of the magnetic field vector over 120\degr\ during a relatively short time
period ($\sim$4~hr) around the peak of the total pressure. This feature together
with a strong magnetic field (up to $\sim$20~nT) and local decrease in $n_{p}$ tentatively
suggests the presence of a magnetic cloud embedded within the CIR. However
other essential characteristics of a magnetic cloud, such as abnormal $T_{p}$ depression
\citep{ric95} and bidirectional beams of suprathermal ($\sim$100~eV)
electrons \citep{gos87} were not observed. The pitch angle distribution
of suprathermal electrons on {\sl STEREO} was examined in
data\footnote{\url{stereo.cesr.fr/plots.php}} from the SWEA instrument \citep{sau08}.
Moreover, the magnetic field rotation occurred near the sector boundary
crossing and thus may be related to the complex magnetic morphology in
the heliospheric current sheet \citep{kle80}. Note that the purported
magnetic cloud is not included in the comprehensive list of the {\sl STEREO}
interplanetary coronal mass ejections compiled by the previously cited team at UCLA.
Recently, \citet{gom11} presented several cases of similar cloud-like
small structures embedded within the CIR which probably favored the formation of reverse
shocks, resulting in enhanced ion acceleration.

\subsubsection{Association with the CIR} \label{ac}

The observations in Figure~\ref{fig4} indicate that ion intensities and $^{3}$He counts on
July 16 commence to increase near the stream interface and show a maximum in the vicinity
of the CIR trailing edge. Such ion intensity increases are commonly
observed in ion intensity-time profiles in CIR events at 1~AU
\citep[e.g.,][]{ric84,mas97,cho00}. We emphasize here that the ion intensities
tend to peak simultaneously at all energies.
Particularly, the low energy channels show a pronounced peak at the trailing boundary
which indicates that the energy spectrum was locally modified. In the standard
basic picture of CIR acceleration, particles can be pre-heated inside
the compression region (e.g., by a stochastic mechanism) and then accelerated
in the reverse shock formed at the CIR trailing boundary. The stream interface
is usually assumed to be a tangential discontinuity through which particle
transport is not expected to occur. Therefore the ion intensity decreases towards
the interface in the CIRs \citep[and references therein]{cro99}. Weak,
sunward streaming of ions with energy $>$1~MeV\,nucleon$^{-1}$ has been reported in CIR events
at 1~AU \citep{ric93}, indicating their source is in the outer heliosphere,
while low energy ions ($<$300~keV) showed anti-sunward streaming in the compressed
solar wind inside CIRs \citep{ric84}.

We can see in Figure~\ref{fig4}d that the high speed solar wind is quite variable behind
the interface. An interesting feature of the high speed wind in the middle of July 17 is
the weaker helium enhancement, mainly seen in the $^{4}$He isotope. The enhancement
commences when there is a sharp drop in the speed and another deflection of
the flow toward the east, while the pressure stays at the background level.
This has similarities to the L1 observations on July 10 in the high-speed
solar wind. $^{4}$He intensity variations coincident with increases or
decreases in solar wind speed have been previously noticed by \citet{mas09}
in the CIR events. The variations have been attributed to the changes in
connection of the spacecraft to different locations in the CIR outside 1~AU.
We find that the 137~keV\,nucleon$^{-1}$ Fe/O abundance ratio is 0.20$\pm$0.07 during the helium
enhancement on July 17 between 08 and 16~UT. Note that statistically significant
Fe/O ratios at higher energies were not available in this time interval. While the helium isotope ratio
($^{3}$He/$^{4}$He) remains enhanced during this interval (as implied by the
mass spectrogram) the Fe/O is smaller than the typical ratio in $^{3}$He-rich
events \citep[$\sim$0.95 at 385~keV\,nucleon$^{-1}$ in][]{mas04}, but somewhat above the
value typical for CIR events \citep[$\sim$0.08 at 150~keV\,nucleon$^{-1}$ in][]{mas97} or
\citep[$\sim$0.10 at 137~keV\,nucleon$^{-1}$ in][]{buc12} suggesting that a mixed population
co-existed during this time interval in the high-speed solar wind.

Further, Figure~\ref{fig4}b shows a brief $^{4}$He increase at the beginning of July 16 which
occurred inside the compression region preceding the stream interface. Ion
intensity enhancements in this part of CIRs have been related to the CIR
forward shock acceleration \citep{bar76}. The enhancements were less
intense or even not present at 1~AU with larger H/He ratio ($\sim$43)
\citep{ric93} than that associated with the CIR trailing edge
\citep[e.g., $\sim$12.5 at 386~keV\,nucleon$^{-1}$ in the survey of][]{buc12}.
The 386~keV\,nucleon$^{-1}$ H/He ratio integrated through this
$^{4}$He increase between 23~UT on July 15 and 04~UT
on July 16 is 45.4$\pm$8.1 and it is strikingly similar to the H/He
ratio reported by \citet{ric93} for corotating increases in the
slow solar wind. The heavy-ion count rates during this enhancement were weak
with the Ne-S and Fe fluences statistically not significant (with errors
greater than 0.5) in the SIT energy range.

%% In this section, we use  the \subsection command to set off
%% a subsection.  \footnote is used to insert a footnote to the text.


%% This section also includes several of the displayed math environments
%% mentioned in the Author Guide.

\section{SOLAR SOURCES} \label{ss}

To determine the solar origin of the $^{3}$He-rich SEPs described in the previous
section we identify the s/c magnetic connection location on the Sun and then examine
flaring at this location using the EUV solar images. For the events which are
associated with energetic electrons and X-ray flares the identification of the responsible
solar source is more straightforward compared to the cases which lack such
an association. In these situations we also use the radio data which generally
show wider longitude coverage than their parent electrons. \citet{nit06} have found
a $\sim$95$\%$ association of $\sim$2-3~MeV\,nucleon$^{-1}$ $^{3}$He-rich SEP events
with type III bursts and only a $\sim$62$\%$ association with $>$30~keV solar electron events.

\subsection{Spacecraft Magnetic Connections} \label{smc}

Figure~\ref{fig5} shows photospheric magnetic field maps with potential-field
source-surface (PFSS) \citep{sch69} model coronal field lines
at the start times of the {\sl STEREO-B} 2011 July 1, {\sl ACE} July 7 and July 9, and
{\sl STEREO-A} July 16 $^{3}$He-rich events. Black and white areas indicate regions
with strong magnetic fields of negative and positive polarity. The regions with predominantly
one polarity correspond to coronal holes and the bipolar areas to active
regions. The field lines shown are those which are open to the heliosphere for
the range of the s/c heliolatitudes. The PFSS extrapolations we use are based
on {\sl Solar Dynamic Observatory} Helioseismic and Magnetic Imager ({\sl SDO}/HMI) magnetograms
which are assimilated into the flux-dispersal model to provide the magnetic field
on the full solar sphere \citep{sch03}. These photospheric
magnetic field maps with a temporal resolution of 6 hours are provided by Lockheed
Martin Solar and Astrophysics Laboratory and are available via the Solar Soft
PFSS package\footnote{\url{www.lmsal.com/$\sim$derosa/pfsspack}}. The black vertical
lines in Figure~\ref{fig5} indicate the east (top panel) and west
(remaining panels) solar limbs as viewed from the Earth. Blue diamonds mark
the footpoints of {\sl STEREO-B} (top panel), L1 (two middle panels) and {\sl
STEREO-A} (bottom panel) at the source surface at 2.5~$R_{\sun}$ from Sun center. The source
surface is a boundary which separates open and closed fields. Outside the source
surface we assume the field follows the Parker spiral. The s/c footpoint
longitudes were determined from the Parker angle using the measured solar wind speed.
Yellow circles indicate the s/c connection location on the Sun. The location
on the Sun was identified by tracking the s/c footpoints over the $^{3}$He-rich period.
The technique which combines the Parker spiral for the interplanetary space and
the PFSS model for the corona has been previously used in identification of
$^{3}$He-rich SEP events sources \citep[e.g.,][]{nit06,wan06}.
In this study we also check whether the in-situ magnetic field polarity matches
the polarity from the PFSS extrapolations.

The top panel in Figure~\ref{fig5} shows that the {\sl STEREO-B} footpoint at the start time
of the 2011 July 1 $^{3}$He-rich period was located near the positive (green)
field lines of the low-latitude coronal hole and $\sim$5\degr--10\degr\ away from the
negative (red) field lines emanating from AR 11243 and AR 11244. It was pointed
out that the footpoint longitude, based on a simple Parker spiral with a constant
solar wind speed, can only be estimated within $\pm$10\degr\
\citep{nol73,kle08}. The in-situ
polarity (see Figure~\ref{fig1}) suggests that the main connection was via negative
field, although a few short excursions to the positive field near the beginning
of the event are also present. Since the intersection points of the negative
polarity field lines from the two ARs, 11243 and 11244, have equal distance to
the {\sl STEREO-B} footpoint, it is not possible to unambiguously determine the connection
region on the Sun. The second panel in Figure~\ref{fig5} shows the L1 footpoint at
the start time of 2011 July 7 $^{3}$He-rich event was connected via negative (red)
field lines to AR 11244. At that time the Earth was generally in the negative
polarity sector with some deviations from the nominal Parker field direction
as indicated by the yellow coding in the polarity color bar in Figure~\ref{fig3}. One
can see in the third panel of Figure~\ref{fig5} that at the start time of 2011 July 9
event the L1 footpoint on the source surface was between the positive (green) field lines of
the low-latitude coronal hole (with nearby AR 11246) and negative (red) field
lines of AR 11243. The in-situ polarity in Figure~\ref{fig3} suggests connection via
positive field. The bottom panel in Figure~\ref{fig5} suggests that {\sl STEREO-A} was connected to
the same coronal hole during the 2011 July 16 $^{3}$He-rich period, in agreement
with the in-situ polarity observations. The coronal hole was presumably the source of
the high-speed solar wind responsible for the CIRs observed on July 9 by
near-Earth s/c and on July 16 by {\sl STEREO-A}. Furthermore, the sequence of the
photospheric maps with a modeled coronal field showed that both July 9 and
July 16 $^{3}$He-rich periods terminated when the s/c lost the
connection to the coronal hole.

\subsection{EUV Observations} \label{eo}

Below we examine data from the Sun Earth Connection Coronal and Heliospheric
Investigation (SECCHI) EUV imager \citep[EUVI;][]{how08} on {\sl STEREO-A} and
the Atmospheric Imaging Assembly \citep[AIA;][]{lem12} on {\sl SDO}. The {\sl STEREO}/EUVI
is taking full-Sun images with 5-minute regular cadence in four EUV wavelengths
and {\sl SDO}/AIA with an unprecedented 12-sec cadence in seven EUV wavelengths. Note
that while {\sl STEREO} is on a heliocentric orbit and increases its separation from
the Earth, {\sl SDO} is in orbit around the Earth

\subsubsection{STEREO-B 2011 July 1 Event} \label{bev}

We have already shown that the 2011 July 1 $^{3}$He-rich period was preceded by
an electron event with energies below $\sim$50~keV. The source-surface model of
the coronal field along with the in-situ polarity suggests connection to the
both ARs 11243 and 11244. The associated X-ray and H$\alpha$ flares are assigned to
the AR 11244 in the NOAA/SWPC list, favoring this region as the most probable
source candidate for the $^{3}$He event. This is further confirmed with the EUV
solar images obtained from {\sl SDO}/AIA. These images reveal eruptions of
material into the corona from AR 11244 starting at $\sim$12:20~UT
with marked brightening around the time of the X-ray flare, but no activity
in AR 11243. \citet{kah87} for instance associated the sources of $^{3}$He-rich
events with filament eruptions. AR 11244 remained active for some time, producing for
example several X-ray flares (mainly B class) on 2011 July 3, as reported
in the NOAA list.

\subsubsection{ACE 2011 July 7 Event} \label{7ev}

The 2011 July 7 $^{3}$He-rich event was associated with two solar electron events
whose associated B7.6 and B6.4 X-ray flares were not assigned to an AR in
the NOAA/SWPC list. Our suggested magnetic connection region on the Sun is
AR 11244. As shown in Figure~\ref{fig5} (2011-Jul-7.9 panel) the region was near the west
solar limb as seen from Earth. The flares might have been partly occulted and
perhaps were stronger than measured.

Figure~\ref{fig6} shows the EUV 131~{\AA} {\sl SDO}/AIA four-minute difference images near the times
of B7.6 (upper panel) and B6.4 (lower panel) flares on 2011 July 7. The displayed
images are the difference of intensity two minutes after and two minutes before the X-ray flare
start times. In both cases the brightening on the west limb was accompanied by eruptions
of the material into the corona nicely visible in the 304~{\AA} {\sl SDO}/AIA channel.
To be sure that the brightenings originated in AR 11244 we examined
{\sl STEREO-A}/SECCHI EUV images. In the {\sl STEREO-A} field of view, AR 11244 was seen
near the central meridian. The eruptions in the AR were, in both cases,
bright for about 20 minutes. The AIA EUV image in the 131~{\AA} channel
corresponds to $\sim$0.4~MK (\ion{Fe}{8}) and to 10~MK (\ion{Fe}{21}) plasma emissions
\citep{lem12}. Previously the 195~{\AA} channel ($\sim$1.6~MK \ion{Fe}{12} and
$\sim$20~MK \ion{Fe}{24}) has been used to identify $^{3}$He-rich SEPs sources
\citep{nit06,wan06}, but we found the brightening on the west limb
easier to distinguish from the bright surroundings in the 131~{\AA} EUV images.
One interesting feature seen in the EUV 193~{\AA} images was the presence of the
large scale loops across the whole AR 11243 while such topology was not obvious
in AR 11244. This is consistent with the PFSS model showing no open field
in AR 11243 (see 2011-Jul-7.9 panel in Figure~\ref{fig5}).

Although it is not clear whether the early 2011 July 8 electron event was
accompanied by $^{3}$He emission (due to the data gap and elevated intensity from
the previous activity) we check the solar source for these electrons. The associated
B3.0 (02:53~UT) X-ray flare is assigned to AR 11243 in the NOAA/SWPC list.
However, with the {\sl STEREO-A} EUV images we see a quite impressive eruption in
AR 11244 at the time of the B3.0 flare and only minor brightening in AR 11243.
AR 11244 was $\sim$2\degr\ behind the west solar limb, but the expulsion was also well
visible in the EUV images on {\sl SDO}. The configuration of the modeled coronal
fields shown in panel 2011-Jul-7.9 in Figure~\ref{fig5} would be valid also for the early
2011 July 8 electron event suggesting a most probable connection to AR 11244.
Furthermore, we note from Figure~\ref{fig3}a that the onset of this electron event is slower
compared to the preceding electron events, which may be caused by the longer
path length for electron propagation as AR 11244 rotated further away. We conclude
that multiple electron events in Figure~\ref{fig3}a, seen within a span of less than one
day have the same solar origin - AR 11244. In the earlier surveys, multiple
electron events within one day often originated in the same active region and
have been often accompanied by a $>$1~MeV\,nucleon$^{-1}$ $^{3}$He-rich increase, but separate
$^{3}$He increases were rarely identified \citep{rea85}. {\sl STEREO-A} EUVI showed
ongoing activity in AR 11244 on July 8 with multiple, smaller compared to the B3.0
flare, brightenings particularly between 10:00 and 11:00~UT, and at 21:15~UT.
No type III bursts were recorded by WAVES instruments
on {\sl WIND} or {\sl STEREO-A} s/c \citep{bou95,bou08} in association with these
brightenings. Also no associated electron events were detected by the near Earth s/c
as seen in Figure~\ref{fig3}a. A notable feature of AR 11244 was multiple
surge-like ejections lasting almost a whole day, clearly visible on the west limb in
AIA 304~{\AA} EUV images.

\subsubsection{ACE 2011 July 9 Event} \label{9ev}

The low-latitude coronal hole, the {\sl ACE} connection area for the 2011 July 9
$^{3}$He-rich event, contained a small recently emerged bipolar active
region, AR 11246, located at 327\degr\ Carrington
longitude. The region was associated with three sunspots on 2011 July 8
as recorded in the SWPC SRS. We searched for activity in that solar region with
particular focus on the period which precedes the beginning of the event.
The {\sl SDO} EUV images show that AR 11246 started to emerge on 2011 July 7
at $\sim$23:00~UT at the western edge of the coronal hole. On 2011 July 8 the AR
was continuously growing, showing two strong brightenings (at 11:40 and 14:40~UT)
before the eruption at 16:30~UT. As viewed also in EUV, the size of AR 11246
($\sim$5$\times$10$^{4}$~km) was much smaller than the neighboring AR 11243 or 11244.
Figure~\ref{fig7} shows the emergence of this region in the 193~{\AA} images from {\sl SDO}. The upper
panel of the figure shows the coronal hole when AR 11246 started to emerge and
the lower panel the AR at the time when it likely stopped expanding. Notice
that coronal holes in EUV images appear as relatively dark regions.
The longitudinal extension of the coronal hole in EUV, between $\sim$20\degr\ and
$\sim$35\degr\ in the upper panel of Figure~\ref{fig7}, agrees well with the open field region from
the PFSS calculation (refer to panel 2011-Jul-7.9 in Figure~\ref{fig5}).

Figure~\ref{fig8} (upper panels) shows AR 11246 around the time of the eruption on 2011 July 8. The figure
is a sequence of three 5-minute difference images from {\sl SDO}/AIA 193~{\AA}.
The middle panel shows collimated or jet-like emissions at the northern footpoint
of the bigger flaring loop (there is likely also a small inner loop) which is
located close to the coronal hole boundary (refer to Figure~\ref{fig7}).
The $\lambda$-shaped structure appears to be consistent with the general scenario of jets
created by reconnection between open and closed field lines \citep{shi92,wan98}.
The lower panel in Figure~\ref{fig8} shows all three ARs 11246, 11243 and 11244 in the
5-minute difference image from {\sl STEREO-A} 195~{\AA} EUVI on 2011 July 8 16:25~UT.
One can see the most prominent brightening in AR 11246 located at E55\degr. Note
that due to projection effects, bright features in AR 11246 appear smaller than
in AR 11244 which is located near the central meridian (W05\degr).
AR 11246 produced several other eruptions on 2011 July 9-10 after the SEP event
start time. On July 11 the AR had rotated close to the west solar limb and was
barely visible with {\sl SDO}. But at that time {\sl STEREO-A} EUVI was able to
see AR 11246 near the central meridian with continuing activity. The AR
decayed in the middle of 2011 July 12, but it still produced sporadic
jets on 2011 July 13-14 before the region re-emerged with an even smaller size at
the end of July 14.

\subsubsection{STEREO-A 2011 July 16 Event} \label{aev}

It has been shown that {\sl STEREO-A} was connected to the same low-latitude coronal
hole during the 2011 July 16 $^{3}$He-rich period. In contrast to the previous case
there were two new bright solar regions in the vicinity of the coronal hole,
one relatively small region beneath the coronal hole southern boundary
(hereafter A1) and a larger one (hereafter A2) $\sim$10\degr\ west of AR 11246. The region
A1 started to emerge on 2011 July 11 and A2 on 2011 July 10. When it started
to emerge the larger region A2 was at $\sim$N08\degr W80\degr\ as viewed from the Earth and
reached its full size behind the solar west limb. Therefore the region does
not clearly show up in the photospheric maps and the PFFS extrapolations probably
could not properly determine the open field region. The limitations of the PFSS
model for unobserved areas especially above rapidly evolving active regions
have been discussed by \citet{ndr08}.

Figure~\ref{fig9} shows the activity just before the start of the 2011 July 16
$^{3}$He-rich period, showing the larger area west of the coronal hole. The upper
panel displays the {\sl STEREO-A} 195~{\AA} EUV image on 2011 July 15 at 22:25~UT.
The lower panel is the difference between the two EUV images taken at 22:25
and 22:20~UT. In the difference image we can see jet-like emission at
the western edge of the coronal hole and no significant brightening in other
regions. The site of this jet coincides with Carrington longitude of AR 11246.
The coronal hole seen in {\sl STEREO-A} EUVI reveals the presence of the open
field region, as inferred from the model field. The region A2
showed one material eruption and one brightening in the first
half of July 15. Careful inspection of this region in the EUV 195~{\AA} images
reveals a closed magnetic field configuration for most of the time presumably
preventing any energetic particles from escaping. The region A1 showed no
significant brightening in EUV.

In addition to the jet in AR 11246 seen near the end of the day, the region
showed three distinct brightenings earlier on July 15 at 12:40, 14:15 and
18:45~UT. Figure~\ref{fig11} shows these brightenings in a wider area of the western
hemisphere to demonstrate that no other regions were simultaneously active.
Bright patches on the west limb were associated with AR 11244. {\sl STEREO-B} saw
AR 11244 at $\sim$E80\degr, and obtained higher cadence (2.5-min) EUV 195~{\AA} images but
revealed no exceptional activity. From the figure we can see, that brightenings in AR 11246 do not have a collimated
ejection or jet-like structure. It has been previously reported that some sources
of $^{3}$He show only amorphous brightening \citep{wan06}. Those authors
pointed out that a jet could have been missed because of the low time cadence (12-minute)
of the EUV observations available. In our case missed jets would have a duration less
than 5 minutes. It is important to note that during the jets and brightenings
in AR 11246 on July 15, {\sl STEREO-A} was magnetically connected via negative field
to nearby AR 11243.

\subsection{Radio Observations Associated with 2011 July 9 and July 16
Events} \label{ro}

Figure~\ref{fig12} (upper panel) shows the {\sl STEREO-A}/WAVES radio spectrogram for
the second half of 2011 July 15 with several type III bursts. It is
interesting that all three brightenings in AR 11246 on July 15, marked by arrows
in Figure~\ref{fig12}, coincide with prominent type III emissions. This indicates that
electrons are accelerated in this region and escape into the interplanetary
medium along the open field lines. These three bursts are characterized by
similar frequency-time profiles and extend to the local plasma frequency at
1~AU \citep[$\leq$50~kHz e.g.,][]{can03}. Note that the jet at 22:25~UT (not marked
in the radio spectrogram) was associated with a narrow type III burst which does
not extend down to the low frequencies as in the previous three cases implying
that the causative electrons probably did not reach $\sim$1~AU. Temporal association
of type III radio bursts with EUV jets in solar impulsive electron events has
been reported in recent studies \citep{kla11,li11}.
Further exploration of {\sl STEREO-A} radio data shows that there were no significant
type III bursts found on 2011 July 16 before the event start time.

We have also examined the {\sl WIND}/WAVES radio
spectrogram of 2011 July 8. Except for a type III
burst associated with the electron event at the beginning of the day there were
two other significant type III emissions on July 8 with start
times at 11:37 and 16:25~UT as shown in the lower panel in
Figure~\ref{fig12}. The first one is associated with the eastern
hemisphere AR 11247 and C2 class flare in the NOAA/SWPC flare list, and
the second one does not have any source association in the list. However, at
the time of the second type III burst (16:25~UT) the {\sl SDO} EUV images showed
a jet-like eruption in AR 11246 displayed in Figure~\ref{fig8}. Such temporal coincidence
implies their common origin and indicates energetic electrons escaping from
AR 11246 into interplanetary space. We remark that the L1 s/c were in
a negative magnetic polarity sector at the time of jet-like eruption and thus
not connected to AR 11246 at the periphery of the coronal hole. We note
another brightening mentioned in Section~\ref{9ev} in {\sl SDO}/AIA EUV observations
in AR 11246 on July 8 just close to the time of the type III burst at 11:37~UT.
This may imply a false association of this type III burst with AR 11247 in
the SWPC list.

\section{SUMMARY AND DISCUSSION} \label{sad}

We have examined $^{3}$He-rich periods of SEPs observed consecutively by the {\sl STEREO-B},
{\sl ACE}, and {\sl STEREO-A} spacecraft when they were widely separated in longitude.
The period observed by {\sl ACE} consists of two distinct $^{3}$He-rich events. The period
observed by {\sl STEREO-B} on 2011 July 1 and later by {\sl ACE} on 2011 July 7 was associated
with a sizeable active region, AR 11244. The AR produced energetic electron events with
associated type III bursts, soft X-ray flares and a H$\alpha$-flare. The period is
characterized by a moderate $^{3}$He-enrichment ($^{3}$He/$^{4}$He $<$1). The period observed
by {\sl ACE} on 2011 July 9 and later by {\sl STEREO-A} on 2011 July 16 was presumably associated
with the small, compact AR 11246 located at the border of the coronal hole and
showed very high $^{3}$He-enrichment and fluence, the highest so far detected on
{\sl STEREO}. The source region produced EUV jets correlated with type III bursts
and exhibited highly dynamic behavior. For example, when the active region
temporarily disappeared the site continued with jet-like emissions.
The characteristics of the events in this study are consistent with earlier
suggestions that small flares have probably more favorable conditions for
the $^{3}$He enrichments of SEPs than large flares \citep{rea88}, and that different size or
morphology of the source active region with diverse flaring could produce
a $^{3}$He-rich event \citep{kah87}.

\begin{figure}
\epsscale{.95}
\plotone{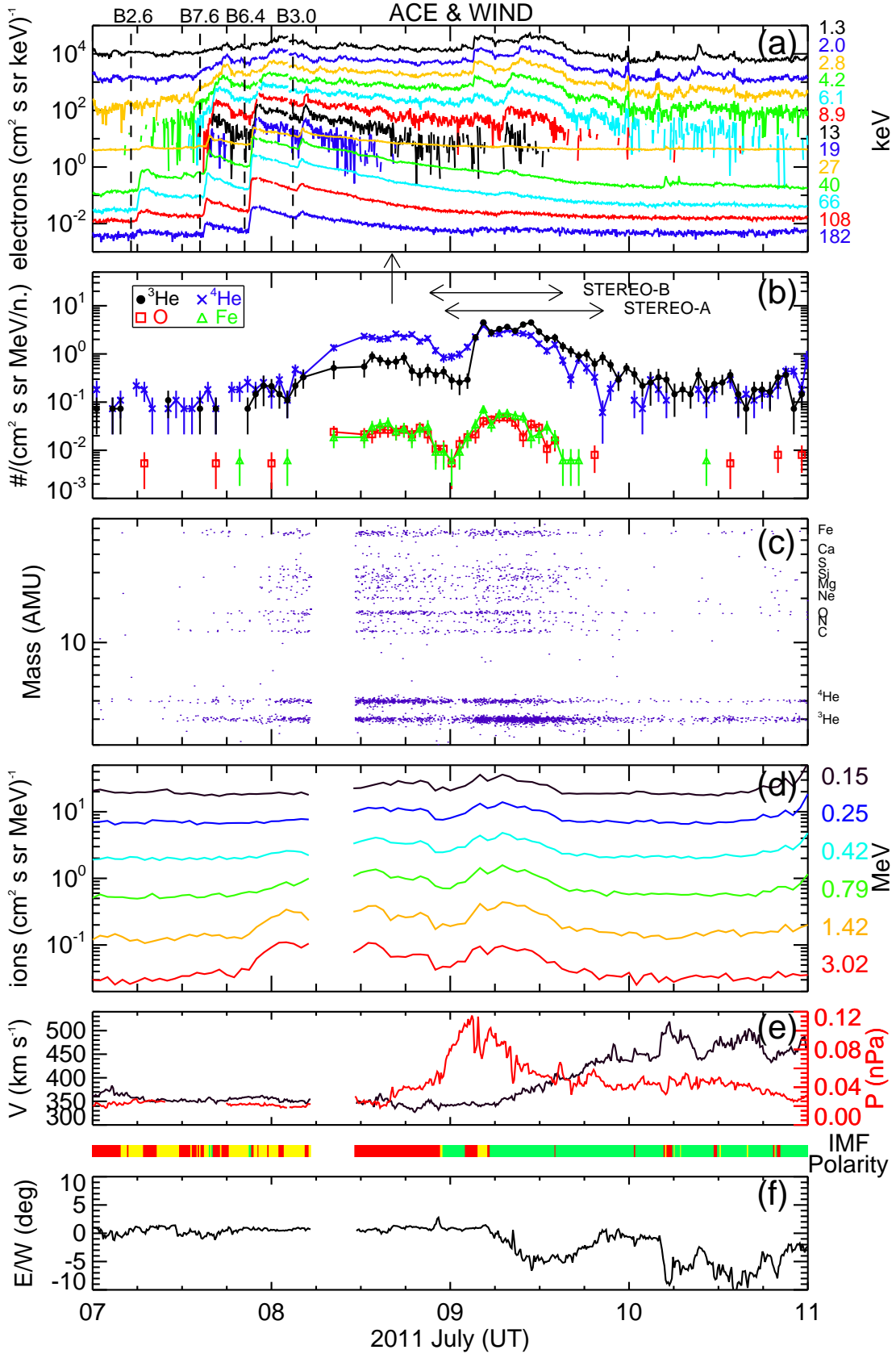}
\caption{(a) 5-min {\sl WIND}/3DP electron intensities from EESA-H (1.3-19~keV)
and SST (27-182~keV) telescopes. Dashed vertical
lines indicate start times of the related {\sl GOES} soft X-ray flares. The arrow
indicates an EUV jet in AR 11246. (b) 1-hr
{\sl ACE}/ULEIS 0.23-0.32~MeV\,nucleon$^{-1}$ $^{4}$He, $^{3}$He, O, Fe
intensity. (c) 0.4-10~MeV\,nucleon$^{-1}$ {\sl ACE}/ULEIS
mass spectrogram. (d) 1-hr {\sl ACE}/EPAM (0.15-3.02~MeV) ion intensities from LEMS120
detector. (e) 10-min solar wind speed $V$ (black curve) and 10-min
total pressure $P$ (red curve). (f) 10-min azimuthal solar wind flow angle in
Radial-Tangential-Normal (RTN) coordinates. Positive angles correspond to flow
in the direction of solar rotation (westward). IMF polarity color bar has
the same meaning as in Figure~\ref{fig1}. \label{fig3}}
\end{figure}

\begin{figure}
%\epsscale{.6}
\plotone{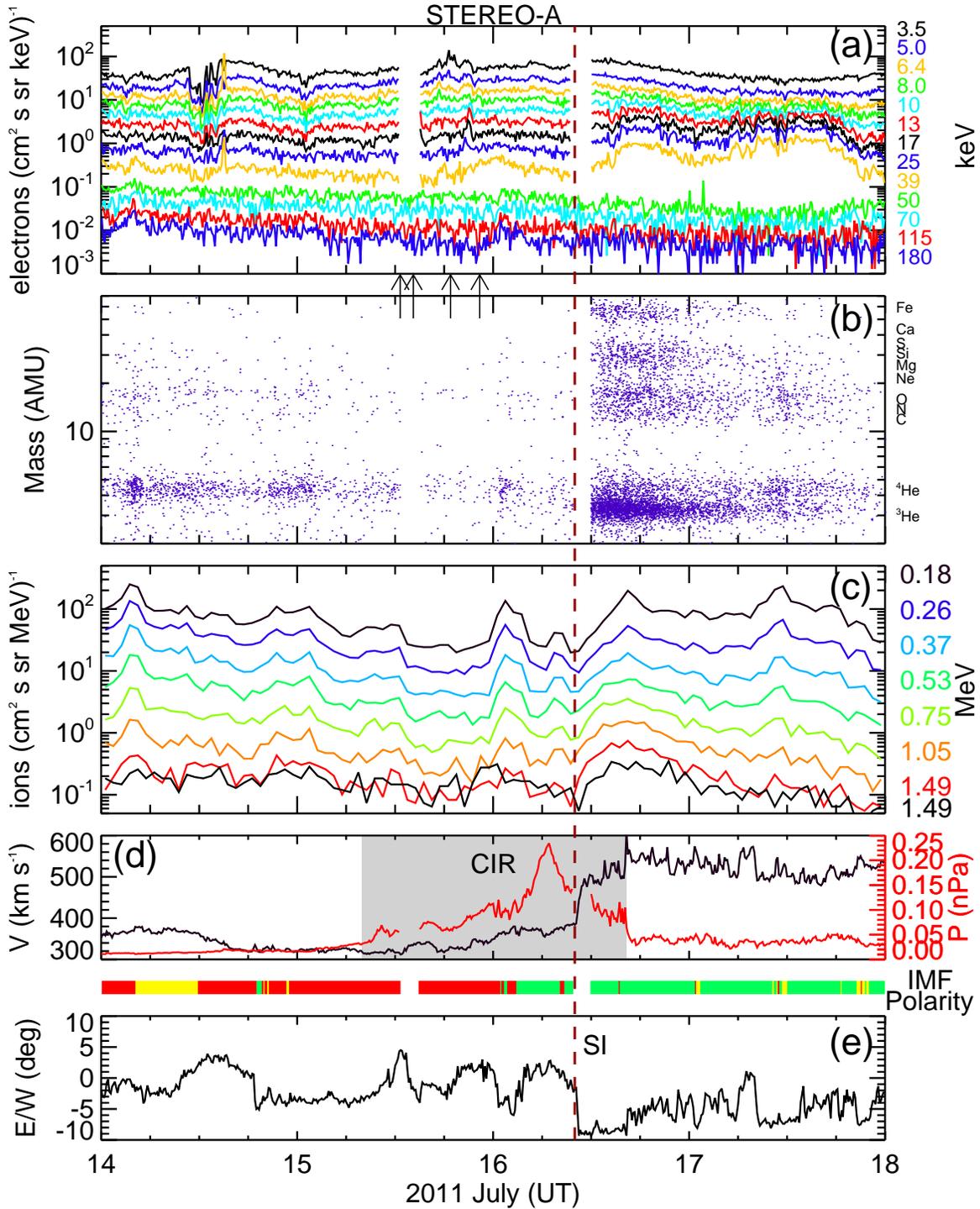}
\caption{(a) 10-min electron intensities from {\sl STEREO-A}/STE D3 (3.5-39~keV)
anti-sunward and {\sl STEREO-A}/SEPT (50-180~keV) sunward
pointing sensors. Arrows indicate EUV brightenings and a jet in AR 11246.
(b) {\sl STEREO-A}/SIT mass spectrogram as in Figure~\ref{fig1}b.
(c) 1-hr {\sl STEREO-A}/SEPT (0.18-1.49~MeV) ion intensities from the sunward pointing
sensor and from the anti-sunward sensor for one energy channel centered at 1.49~MeV
(lower black curve). (d) 10-min solar wind speed $V$ (black curve) and 10-min total pressure $P$
(red curve). Grey shaded region marks the time interval of the CIR. Dashed
vertical line indicates solar wind stream interface (SI). (e) 10-min azimuthal
solar wind flow angle in RTN coordinates. IMF polarity color bar has the
same meaning as in Figure~\ref{fig1}. \label{fig4}}
\end{figure}

\begin{figure}
\epsscale{.8}
\plotone{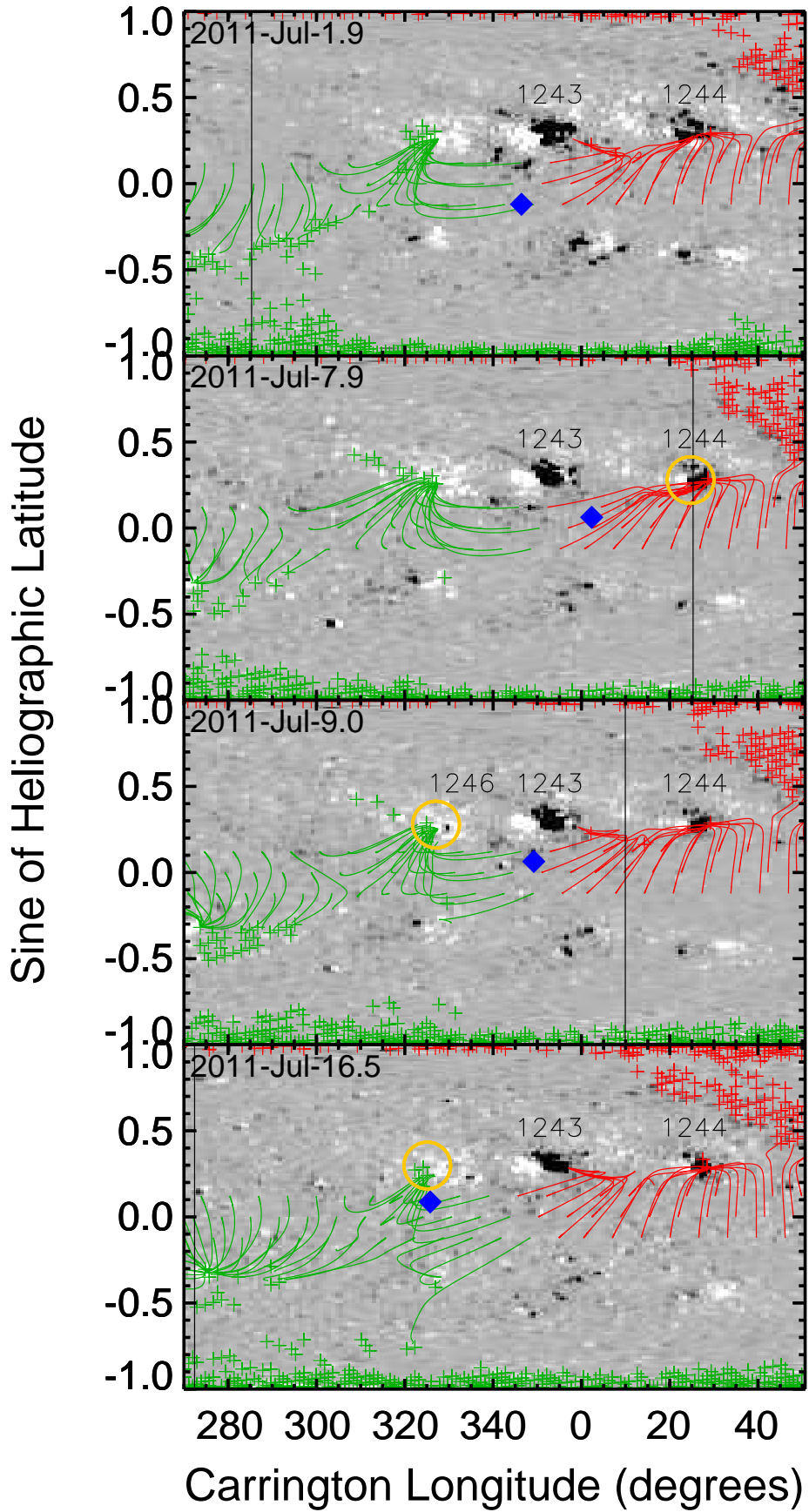}
\caption{Photospheric magnetic field scaled to $\pm30$~Mx\,cm$^{-2}$
(gray scale) with PFSS model coronal field lines (red and green)
at the start time (as date and fraction of day) of the $^{3}$He-rich events. Shown are field
lines which intersect source surface at latitudes 0\degr\ and $\pm$7\degr. Red/green
indicates negative/positive open field. Blue diamonds mark {\sl STEREO-B}
(top panel), L1 (two middle panels) and {\sl STEREO-A} (bottom panel) footpoints on
the source surface. Black vertical lines mark east (top panel) and west
(three lower panels) solar limb from the Earth view point. Yellow circles
(three lower panels) indicate the s/c connection location on the Sun. NOAA
ARs 11243, 11244 and 11246 are indicated. Note that time runs from right to
left in the maps. \label{fig5}}
\end{figure}

\begin{figure}
\plotone{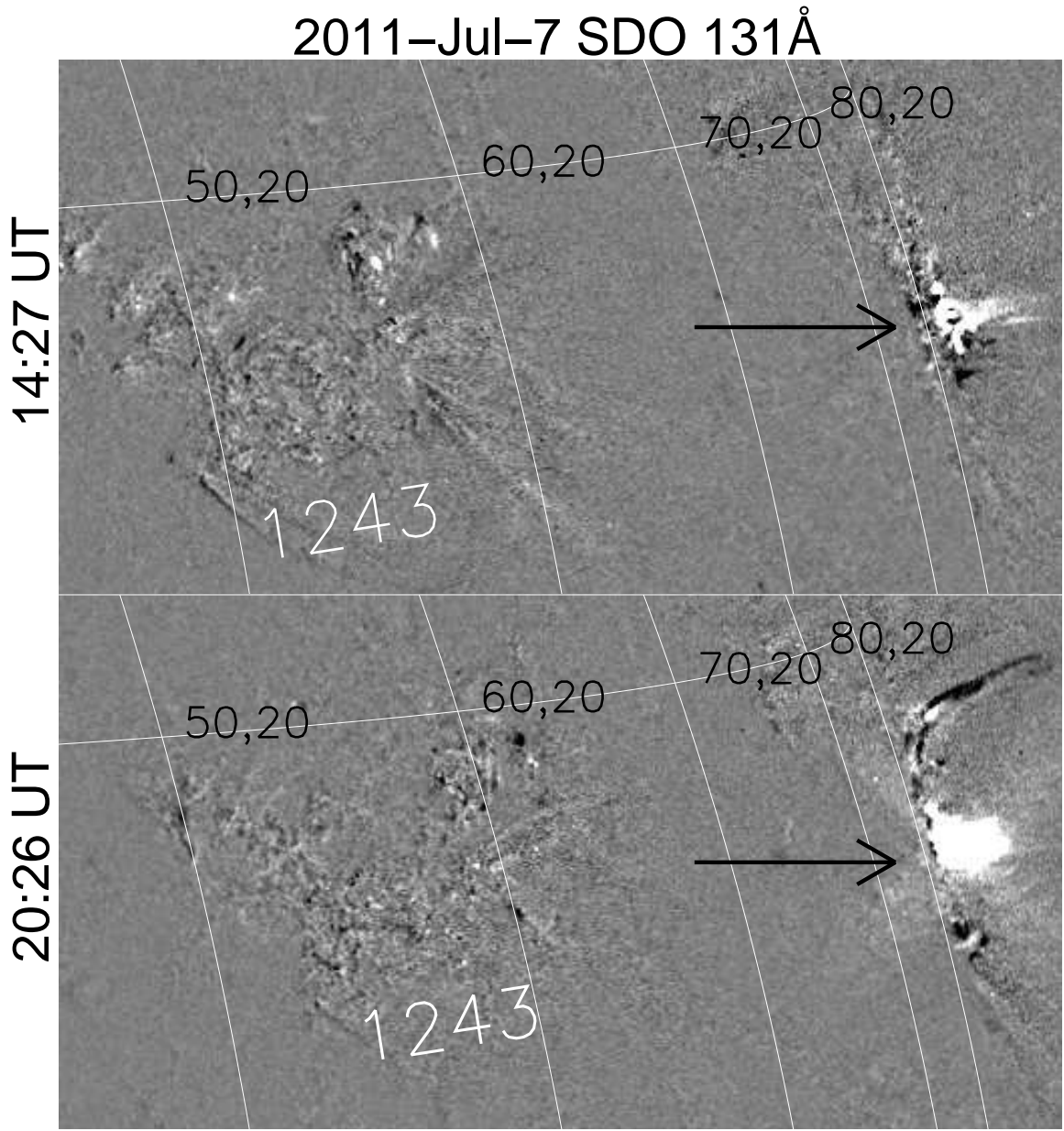}
\caption{EUV 131 {\AA} {\sl SDO}/AIA four-minute difference images near the times of
B7.6 (upper panel) and B6.4 (lower panel) X-ray flares on 2011 July 7; white
(black) indicates increasing (decreasing) emission over the difference time interval.
Black arrows point to the brightenings in AR 11244. AR 11243 and the heliographic
longitude - latitude grid are labeled. \label{fig6}}
\end{figure}

\begin{figure}
\plotone{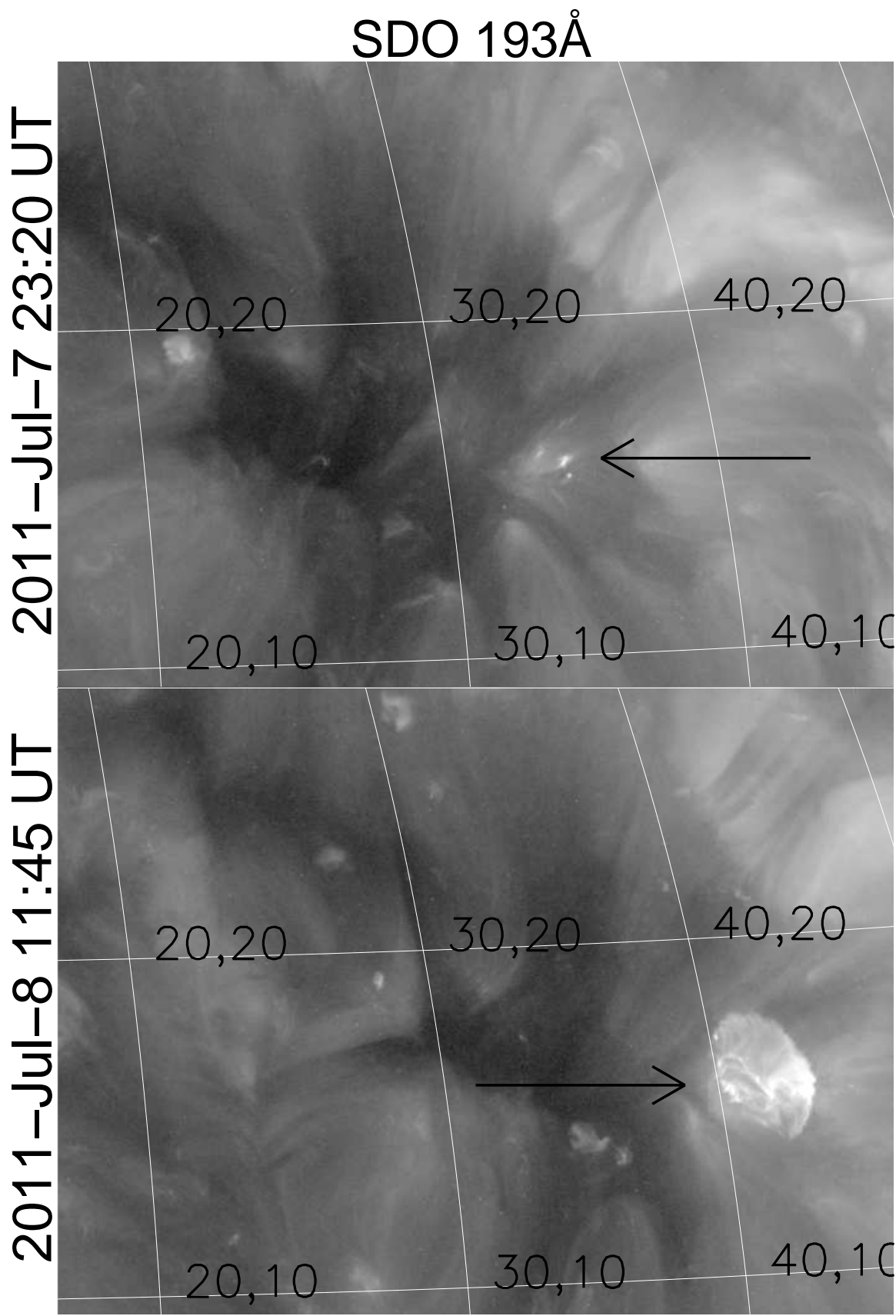}
\caption{EUV 193 {\AA} {\sl SDO}/AIA images of the coronal hole on 2011 July 7 at
23:20~UT (upper panel) and on 2011 July 8 at 11:45~UT (lower panel). Black
arrow marks an emergence of AR 11246. Heliographic longitude - latitude grid
with labels in degrees is displayed. \label{fig7}}
\end{figure}

\begin{figure}
%\epsscale{.40}
%\plotone{f8r.eps}
\includegraphics[angle=90,scale=0.75]{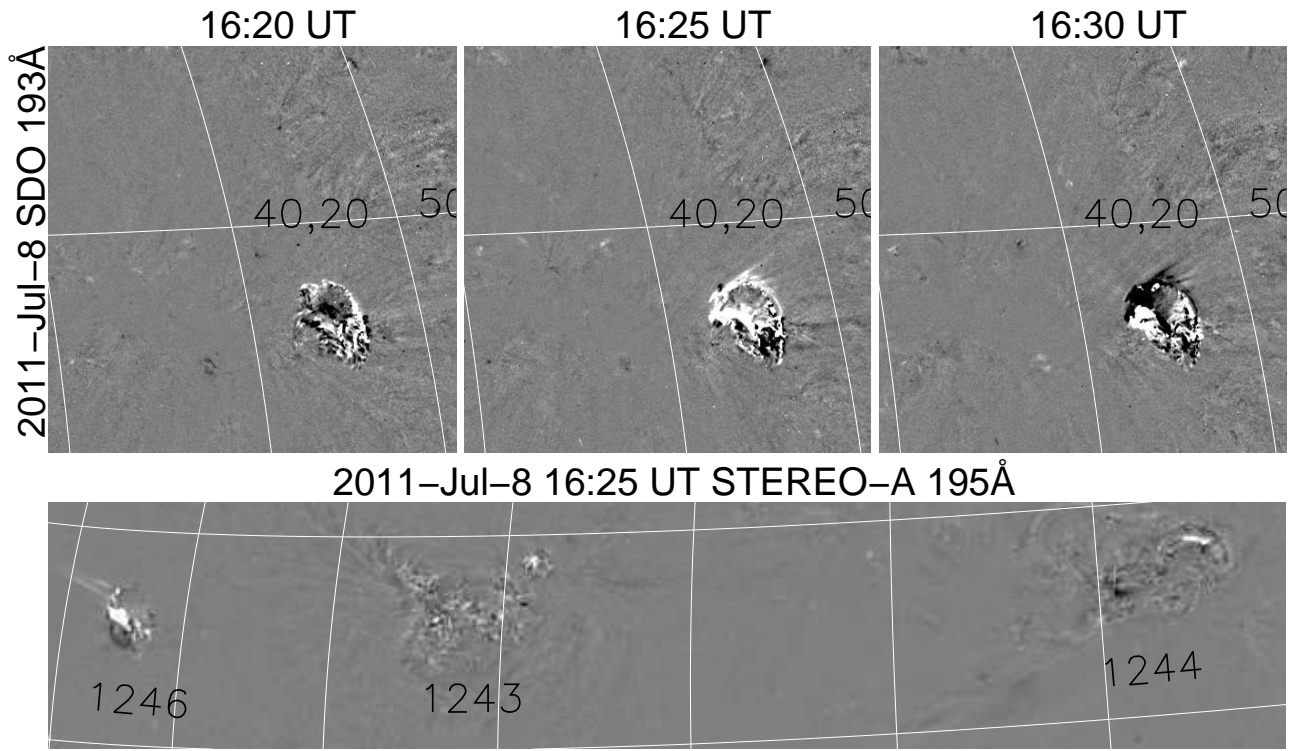}
\caption{5-minute difference EUV 193~{\AA} {\sl SDO}/AIA (upper panels) and
195~{\AA} {\sl STEREO-A}/SECCHI (lower panel) images on 2011 July 8 near the
eruption time of AR 11246. \label{fig8}}
\end{figure}

\begin{figure}
\plotone{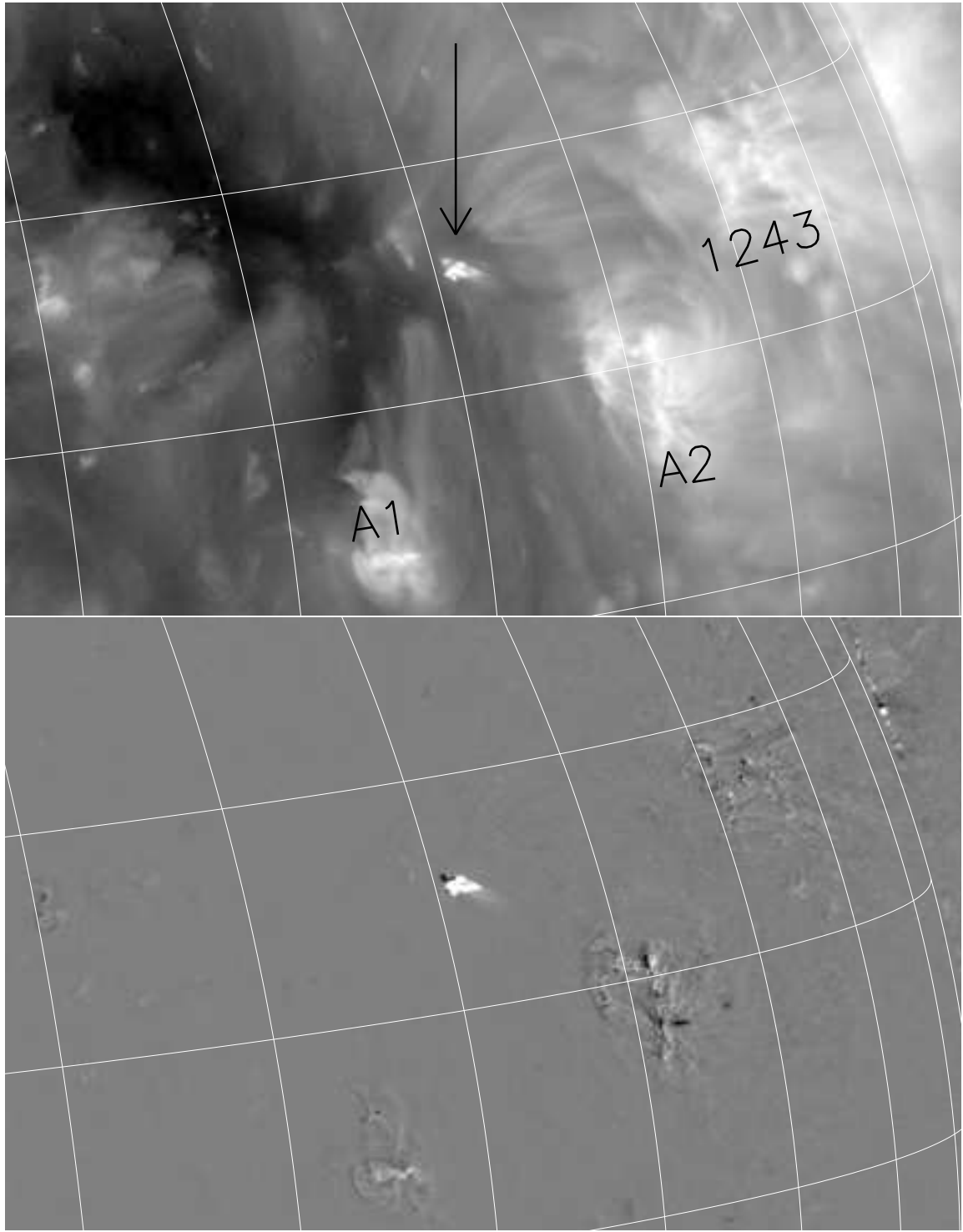}
\caption{{\sl STEREO-A}/SECCHI 195 {\AA} EUV regular (upper panel) and five-minute
difference (lower panel) image on 2011 July 15 at 22:25~UT. Marked are regions
A1, A2, and AR 11243. The heliographic longitude - latitude grid has 10\degr\ spacing.
The arrow points to the jet in AR 11246 located at coronal hole border
at $\sim$N15\degr W40\degr. \label{fig9}}
\end{figure}

\begin{figure}
\epsscale{.65}
\plotone{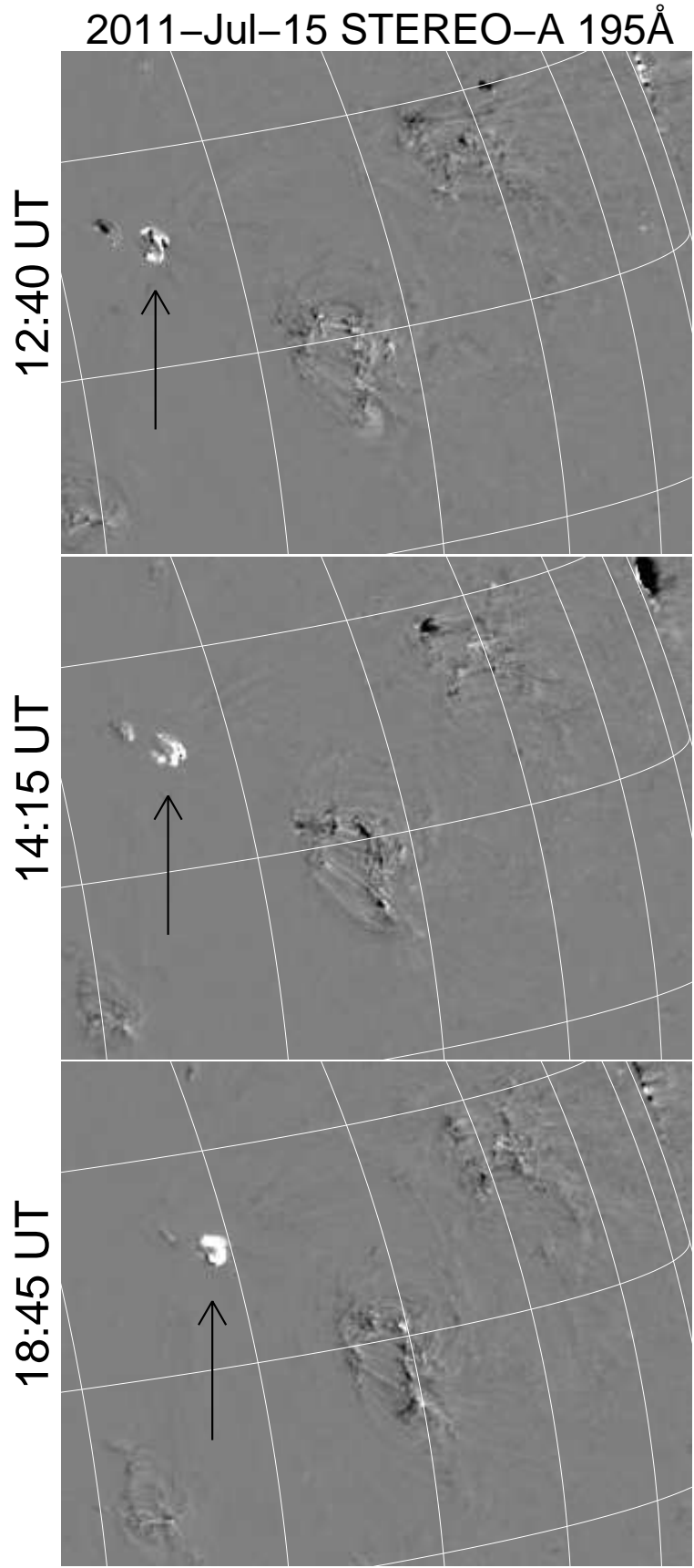}
\caption{{\sl STEREO-A}/SECCHI 195 {\AA} EUV 5-minute difference images on 2011 July 15
at 12:40, 14:15 and 18:45~UT. Arrows point to the brightening in AR 11246 at
the western coronal hole border.  \label{fig11}}
\end{figure}

\begin{figure}
%\epsscale{.71}
%\plotone{waves_sta1.eps}
\includegraphics[angle=90,scale=0.8]{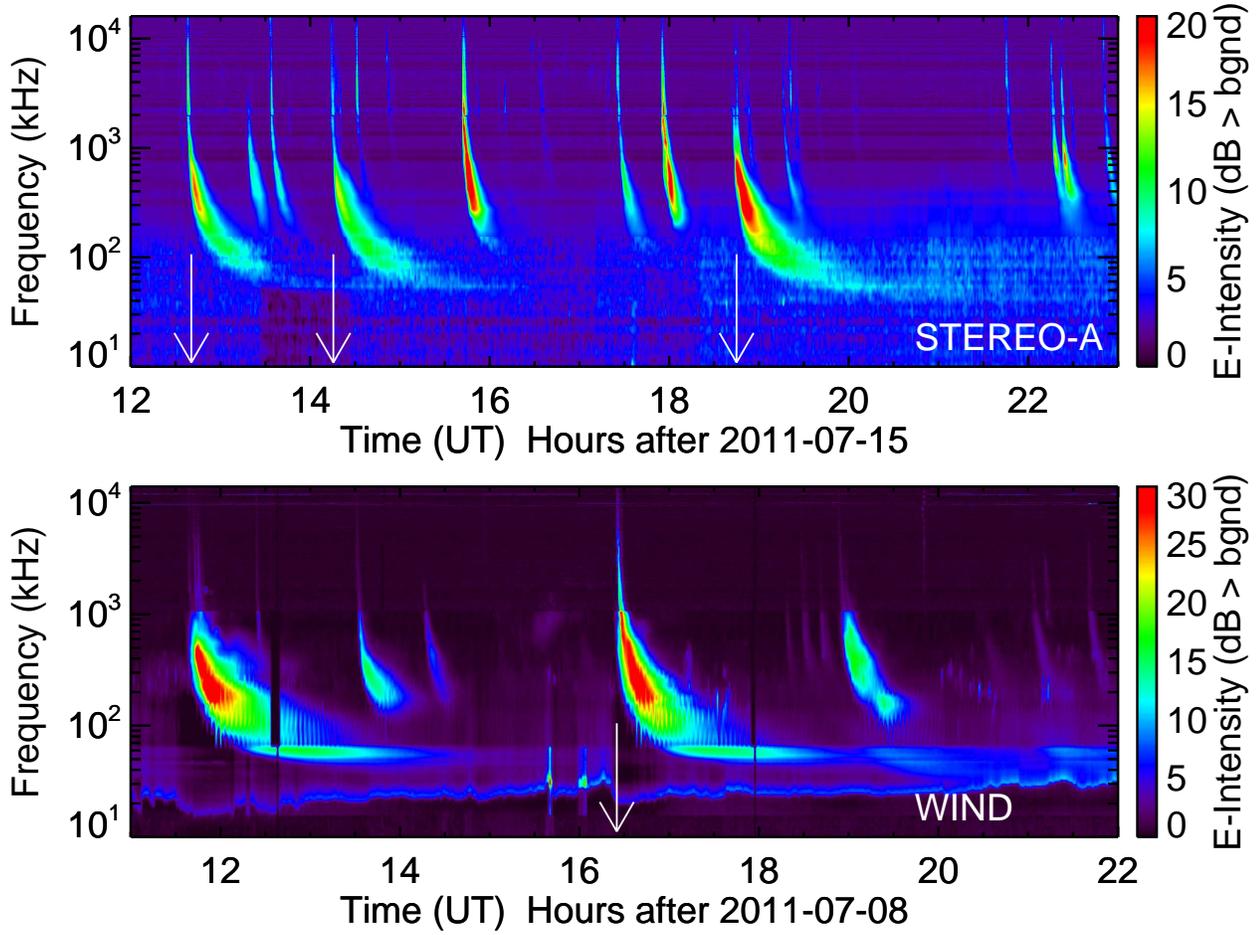}
\caption{Radio spectrogram from {\sl STEREO-A} (upper panel) and WIND
(lower panel) WAVES instruments. Arrows indicate
time of the brightenings and jet in AR 11246. \label{fig12}}
\end{figure}

In spite of a dispersionless onset and absence of energetic electrons in
the July 9 and 16 events the high cadence EUV images combined with the radio
observations allowed us to determine the likely source brightening for
the $^{3}$He emission. The approximate travel time of solar $^{3}$He ions with
an energy 0.23-0.32~MeV\,nucleon$^{-1}$ is about 7 hours to {\sl ACE} along a nominal spiral with
length of 1.2~AU. If the $^{3}$He was released during the eruption in AR 11246 at
16:30~UT on 2011 July 8 the ions with that energy should have been detected
at the beginning of July 9, which is approximately consistent with the start
of the event in the low energy time-intensity profile in Figure~\ref{fig3}b. Since
the s/c was not connected to the coronal hole (and to the source AR) around
the eruption time it missed the electrons and the first-arriving, higher-energy
ions. This explains the lack of velocity dispersion in the ions received after the s/c enters
the field lines connected to the coronal hole. If the release of the energetic ions
for the 2011 July 16 period was associated for example with the jet and type III
radio burst at 22:25~UT on July 15 then the lowest energy ions (0.18~MeV)
shown in Figure~\ref{fig4}c should arrive at {\sl STEREO-A} around 06:30~UT on
July 16 (i.e., three hours before the dispersionless event onset). If
the ions were released during the brightening at 18:45~UT, which was
associated with the prominent type III burst, the highest energy ions
(1.49~MeV) should have only a $\sim$3~hr delay after the flare. This leads to
the question of where these high energy solar ions, with enormous $^{3}$He enrichment,
were residing in the heliosphere for about half of the day prior to the
observed onset.

The timing of the ion intensities in the 2011 July 16 period indicates that
energetic $^{3}$He is closely associated with the CIR which is further supported
by the pre- and post-event abundances. Such association may arise when a source
active region is located at the periphery of the coronal hole with open field
lines bending to the ecliptic. The high-speed solar wind emanating from the hole
may create a corotating compression region in the heliosphere near $\sim$1 AU.
Thus the energetic $^{3}$He injected from the solar source near the hole is guided
by the open field to the CIR. The propagation of SEPs in the compressed field
in CIRs has been modeled by \citet{koc03}. The authors employed
the \citet{gia02} model of CIR acceleration on the solar
wind speed gradients at 1~AU. Their simulations show that the magnetic
enhancement associated with the CIR presents a kind of magnetic mirror away
from the Sun where solar particles may be temporally trapped and re-accelerated.
Indeed, our observations show a significant sunward component during
the 2011 July 16 $^{3}$He-rich period indicating a reflecting boundary beyond
the observer. \citet{koc08} have reported on multi-day $^{3}$He-rich
periods that essentially all have dispersionless onset and were associated
with compression in the solar wind but such convincing associations
as in the July 16 period have not been presented. Those authors suggest that confinement
of particles in CIRs is a significant factor of extended $^{3}$He-rich periods
of SEPs.

It is not clear if the weaker CIR seen during the 2011 July 9 $^{3}$He-rich period
had a marked influence on the SEPs. The event commenced with a change of the magnetic
connection and not with the start of the solar wind speed rise or at the CIR
stream interface although the later was not fully developed. Presumably at
larger radial distances where the corotating compression regions are found to
be stronger, a closer association might exist. \citet{koc03} concluded
that the effects of the compression may be important if the observed
solar wind speed increase near Earth is more than 100~km\,s$^{-1}$ within a few hours. In the
2011 July 9 CIR, such a speed change is observed during a quite long $\sim$12~hr
period. Notice that no significant concurrent CIR event was observed in association
with the 2011 July 9 and July 16 compression corotating regions. That means
at least for the strong July 16 compression that the bulk solar wind was not
accelerated in the CIR. It is consistent with recent suggestions that CIR ions
are accelerated out of the suprathermal ion pool \citep{mas12}. In
the case of the July 16 CIR the suprathermal population was probably dominated by
the $^{3}$He-rich SEPs which might undergo acceleration in the strong compression
region.

These small SEP events are often difficult to conclusively identify with source
activity in active periods when there may be multiple candidates. In the July 9
event the s/c footpoint lies in a reasonable distance from both ARs 11246 and 11244.
Note that the distance to AR 11244 was still within 3$\sigma$ ($\sim$48\degr) of the flare longitudes
distribution associated with $^{3}$He-rich SEP events derived by \citet{rea99}. There are a few
reasons to choose AR 11246 over AR 11244. The type III emissions at 11:37
and 16:25~UT preceding the July 9 event were probably associated with AR 11246
because of their temporal coincidence with the EUV brightenings in this AR; the brightenings,
fainter at these times in AR 11244 were not new and likely continued from the previous
activity. If the type III burst at 16:25~UT was associated with AR 11244 then
why do we not detect escaping electrons, while being in a negative IMF
sector and presumably connected to AR 11244? Another reason in favor of AR 11246
is the reversal of the IMF to the polarity at AR 11246 at the start of
the July 9 event. Although AR 11244 continued with some activity on July 15-16,
the type III bursts at 12:40, 14:15 and 18:45~UT on July 15 appear
to be correlated with EUV brightenings in AR 11246. Note that some ambiguity may
still remain in locating these type III bursts because we use lower (5 minute)
cadence EUVI data. {\sl WIND}/WAVES observed weak and only low-frequency counterparts
of these bursts. {\sl STEREO-B}/WAVES did not record any type III emission on July 15 in the same
time period as {\sl STEREO-A}. This implies that the source was well hidden
from the {\sl STEREO-B} view, which was not case of AR 11244. In addition,
the AR 11244 location is less likely to be the source for the $^{3}$He-rich
event on {\sl STEREO-A}. We have shown that the
{\sl STEREO-A} footpoint was co-located with the coronal hole (containing AR
11246), while AR 11244 was quite far, 60\degr\ west of the footpoint.

In conclusion, with the advantage of widely separated spacecraft carrying advanced
imaging, radio and particle instrumentation, supported by the modeling of
the coronal field, we have identified solar ARs which exhibited repeated
energetic electron and $^{3}$He emissions for a relatively long interval of time,
up to about a quarter of a solar rotation. This is significantly longer than
single s/c observations which previously reported $^{3}$He injections
from the same AR over a period of one day. Thus this study suggests that
the conditions for the $^{3}$He acceleration in a single solar region may persist
for a long time. Furthermore the observations in this paper show that a recurrent
$^{3}$He-rich period can be preserved in the solar wind by confining SEPs in
the CIR. The observations of a long-lived recurrent period presented in this
study may provide new insights on the acceleration of $^{3}$He ions in the solar
flares and their transport in the heliosphere. In particular, extended $^{3}$He-rich
periods recently discovered with single s/c observations may be naturally
explained by multiple injections from the same AR which may presumably last
over days.

%% The \notetoeditor{TEXT} command allows the author to communicate
%% information to the copy editor.  This information will appear as a
%% footnote on the printed copy for the manuscript style file.  Nothing will
%% appear on the printed copy if the preprint or
%% preprint2 style files are used.


%% Included in this acknowledgments section are examples of the
%% AASTeX hypertext markup commands. Use \url without the optional [HREF]
%% argument when you want to print the url directly in the text. Otherwise,
%% use either \url or \anchor, with the HREF as the first argument and the
%% text to be printed in the second.

\acknowledgments

This work was supported by the Bundesministerium f\"{u}r Wirtschaft through
the Deutsches Zentrum f\"{u}r Luft- und Raumfahrt (DLR) under grant 50 OC 0904.
The work at JHU/APL was supported by NASA under contract SA4889-26309 from
the University of California Berkeley and by NASA grants 44A-1089749 and
NNX13AR20G. R.~G\'{o}mez-Herrero acknowledges the financial support from the Spanish
Ministerio de Ciencia e Innovaci\'{o}n under project AYA2011-29727-C02-01.
The {\sl STEREO}/SEPT project is supported by DLR grant 50 OC 1302. The authors
thank M.~E.~Wiedenbeck for providing the \textsl{STEREO}/LET helium mass spectrogram
plots. We acknowledge the National Space Science Data Center (NSSDC), Space Physics Data Facility
(SPDF) and PI K.~W.~Ogilvie at GSFC and A.~J.~Lazarus at MIT for use of the
{\sl WIND}/SWE data. We thank the {\sl ACE} SWEPAM/EPAM/MAG and the {\sl STEREO}
PLASTIC instrument teams for the use of the solar wind plasma data.

\clearpage

%% the file, uncomment the \plotone, \plottwo, and
%% If you need a layout that cannot be achieved with \plotone or
%% \plottwo, you can invoke the graphicx package directly with the
%% \includegraphics command or use \plotfiddle. For more information,
%% please see the tutorial on "Using Electronic Art with AASTeX" in the
%% documentation section at the AASTeX Web site, http://aastex.aas.org/
%%
%% The examples below also include sample markup for submission of
%% supplemental electronic materials. As always, be sure to check
%% the instructions to authors for the journal you are submitting to
%% for specific submissions guidelines as they vary from

%% This example uses \plotone to include an EPS file scaled to
%% 80% of its natural size with \epsscale. Its caption
%% has been written to indicate that additional figure parts will be
%% available in the electronic journal.

\clearpage

\clearpage

\clearpage

\clearpage

\clearpage

%\clearpage

%\clearpage

%\clearpage

%\clearpage

%\begin{figure}
%%\epsscale{.40}
%%\plotone{aia_193A_2011-07-08T16_25_diff_new.eps}
%\includegraphics[angle=90,scale=0.68]{f10.eps}
%\caption{Temporal evolution of EUV jet in AR 11246 at coronal hole boundary on
%2011 July 15. Each panel presents difference between two {\sl STEREO-A}/SECCHI
%195~{\AA} EUV images taken 5 minutes apart. The heliographic longitude - latitude
%grid has 10\degr\ spacing.  \label{fig10}}
%\end{figure}

%\clearpage

%% Here we use \plottwo to present two versions of the same figure,
%% one in black and white for print the other in RGB color
%% for online presentation. Note that the caption indicates
%% that a color version of the figure will be available online.
%%

%% If you are not including electonic art with your submission, you may
%% mark up your captions using the \figcaption command. See the
%% No more than seven \figcaption commands are allowed per page,
%% so if you have more than seven captions, insert a \clearpage

%% Tables should be submitted one per page, so put a \clearpage before

%% Two options are available to the author for producing tables:  the
%% deluxetable environment provided by the AASTeX package or the LaTeX
%% table environment.  Use of deluxetable is preferred.
%%

%% Three table samples follow, two marked up in the deluxetable environment,
%% one marked up as a LaTeX table.

%% In this first example, note that the \tabletypesize{}
%% command has been used to reduce the font size of the table.
%% We also use the \rotate command to rotate the table to
%% landscape orientation since it is very wide even at the
%% Note also that the \label command needs to be placed

%% This table also includes a table comment indicating that the full
%% version will be available in machine-readable format in the electronic

\clearpage

\begin{deluxetable}{lcccc}
\tabletypesize{\scriptsize}
%\rotate
\tablecaption{$^{3}$He-Rich Period Properties \label{tab1}}
\tablewidth{0pt}
\tablehead{
\colhead{Start Day} & \colhead{2011 Jul 1} & \colhead{2011 Jul 7} &
\colhead{2011 Jul 9\tablenotemark{a}} & \colhead{2011 Jul 16\tablenotemark{a,b}}
}
\startdata
Spacecraft (s/c) & {\sl STEREO-B} & {\sl ACE} & {\sl ACE} & {\sl STEREO-A} \\
s/c Carrington Lon.\tablenotemark{c} & 283\degr & 296\degr & 280\degr & 281\degr \\
s/c Heliographic Lat. & -6.9\degr & 3.6\degr & 3.7\degr & 5.0\degr \\
$^{3}$He-Rich Interval (doy) & 182.9--183.6 & 188.9--190.0 & 190.0--191.0  & 197.5--198.3 \\
$^{3}$He/$^{4}$He\tablenotemark{d} & 0.74$\pm$0.17 & 0.42$\pm$0.04 & 2.35$\pm$0.13 & 4.08$\pm$0.29 \\
Fe/O\tablenotemark{d} & \nodata & 0.91$\pm$0.15 & 1.31$\pm$0.20 & 1.19$\pm$0.31 \\
$^{3}$He Fluence\tablenotemark{d} ($\times10^3$) & 3.73$\pm$0.63 & 20.6$\pm$1.6 & 127$\pm$4  & 96.1$\pm$3.2 %\\
\enddata
%% Text for table notes should follow after the \enddata but before
%% the \end{deluxetable}. Make sure there is at least one \tablenotemark
%% in the table for each \tablenotetext.
\tablenotetext{a}{event included in survey by \citet{buc13a}}
\tablenotetext{b}{event included in survey by \citet{buc13b}}
\tablenotetext{c}{at event start time}
\tablenotetext{d}{320-450~keV\,nucleon$^{-1}$; fluence units - particles (cm$^2$\,sr\,MeV/nucleon)$^{-1}$}
\end{deluxetable}

%% The following command ends your manuscript. LaTeX will ignore any text

\end{document}